\documentclass[iop]{emulateapj}

\usepackage{natbib,aas_macros}
\usepackage{color}
\usepackage{ulem}
\usepackage{dcolumn}
\usepackage{bm}
\usepackage{amsmath}
\usepackage[flushleft]{threeparttable}

\newcommand{\del}{\partial}
\renewcommand{\thefootnote}{\fnsymbol{footnote}}

\renewcommand{\apjl}{ApJL}

\shorttitle{Neutrino-driven Ejecta from NS-NS Merger}
\shortauthors{S.Fujibayashi, Y.Sekiguchi, K.Kiuchi, and M.Shibata}

\begin{document}

\title{Properties of Neutrino-driven Ejecta from the Remnant of Binary Neutron Star Merger : Purely Radiation Hydrodynamics Case}

\author{
Sho~Fujibayashi
}
\affil{
Yukawa Institute for Theoretical Physics, Kyoto University, Kyoto 606-8502, Japan
}
\email{
sho.fujibayashi@yukawa.kyoto-u.ac.jp
}

\author{
Yuichiro~Sekiguchi
}
\affil{
Department of Physics, Toho University, Funabashi, Chiba 274-8510, Japan
}
\author{
Kenta~Kiuchi and Masaru~Shibata
}
\affil{
Center for Gravitational Physics, Yukawa Institute for Theoretical Physics, Kyoto University, Kyoto 606-8502, Japan
}

\keywords{accretion, accretion disks--neutrinos--stars: neutron--gamma-ray burst: general--relativistic processes}

\begin{abstract}
We performed general relativistic, long-term, axisymmetric neutrino radiation hydrodynamics simulations for the remnant formed after the binary neutron star merger, which consist of a massive neutron star and a torus surrounding it.
As an initial condition, we employ the result derived in a three-dimensional, numerical relativity simulation for the binary neutron star merger.
We investigate the properties of neutrino-driven ejecta.
Due to the pair-annihilation heating, the dynamics of the neutrino-driven ejecta is significantly modified.
The kinetic energy of the ejecta is about two times larger than that in the absence of the pair-annihilation heating.
This suggests that the pair-annihilation heating plays an important role in the evolution of the merger remnants.
The relativistic outflow, which is required for driving gamma-ray bursts, is not observed because the specific heating rate around the rotational axis is not sufficiently high due to the baryon loading caused by the neutrino-driven ejecta from the massive neutron star.
We discuss the condition for launching the relativistic outflow and the nucleosynthesis in the ejecta.
\end{abstract}

\section{Introduction} \label{sc:intro}
Binary neutron star (NS-NS) merger is one of the promising sources of gravitational waves for ground-based gravitational-wave detectors, such as advanced LIGO \citep{2010NIMPA.624..223A}, advanced Virgo \citep{2011CQGra..28k4002A}, and KAGRA \citep{2010CQGra..27h4004K}.
In association with gravitational waves, the merger remnant, which is composed typically of a central compact object (NS or black hole) surrounded by a massive accretion torus, emits a huge amount of neutrinos.
Neutrinos emitted could be a major source of various types of phenomena as follows.

First, a large amount of mass could be ejected as sub-relativistic neutrino-driven wind \citep{2009ApJ...690.1681D,2014MNRAS.441.3444M,2014MNRAS.443.3134P}, in addition to the dynamical mass ejection, which would occur at NS-NS mergers.
Then, a substantial amount of radioactive nuclei are likely to be synthesized via the $r$-process \citep{1974ApJ...192L.145L,1982ApL....22..143S,1989Natur.340..126E,1999A&A...341..499R,1999ApJ...525L.121F,2011ApJ...738L..32G,2014ApJ...789L..39W}.
The ejecta is heated by the radioactive decay and fission of these heavy elements and would shine at optical and infra-red bands in $1-10$ days after the merger \citep{1998ApJ...507L..59L,2013ApJ...774...25K,2013ApJ...775..113T}.
Simultaneous detection of this so-called ``kilonova" or ``macronova" with gravitational waves will significantly improve positional accuracy of gravitational wave sources.
The neutrino-driven ejecta could be an additional energy source to the dynamical ejecta.

Second, if a relativistic outflow is launched by neutrino heating, the merger remnant would drive short-duration gamma-ray bursts (SGRBs).
The neutrino pair-annihilation could be the engine for driving the extremely relativistic ejecta \citep{1989Natur.340..126E,1992MNRAS.257P..29M,1992ApJ...395L..83N}.
This is because this process can deposit thermal energy into materials regardless of the baryon density, so that the ejecta could achieve large photon-to-baryon ratio if the baryon density is sufficiently low in the heating region.
We should note that there are several other scenarios proposed to drive the relativistic ejecta \citep{2007PhR...442..166N}, such as the electromagnetic energy extractions from rotating black holes \citep{1977MNRAS.179..433B} and rapidly rotating pulsars of a strong magnetic field \citep{1992Natur.357..472U}.

Radiation hydrodynamics simulations of a merger-remnant torus surrounding a central compact object have been performed by many authors \citep{2009ApJ...690.1681D,2013PhRvL.111j8301F,2014MNRAS.441.3444M,2014MNRAS.443.3134P,2015MNRAS.446..750F,2015MNRAS.448..541J,2016ApJ...816L..30J,2016PhRvD..93d4019F}.
Recently, simulations of black hole-torus system approximately taking the pair-annihilation heating into account were performed \citep{2016ApJ...816L..30J}.
It was concluded that the system is unlikely to drive SGRBs because the neutrino heating cannot overcome the baryon loading.

Recent numerical relativity simulations of the NS-NS mergers with finite-temperature equations of state (EOS) that can support the observed $\sim 2M_\odot$ NSs \citep{2010Natur.467.1081D,2013Sci...340..448A} have suggested that a massive neutron star (MNS) is likely to be formed as the remnant of the merger if the total binary mass is not extremely high ($\lesssim 2.8 M_\odot$) \citep{2005PhRvD..71h4021S,2006PhRvD..73f4027S,2009PhRvD..80f4037K,2010PhRvL.104n1101K,2011PhRvL.107e1102S,2011PhRvD..83l4008H,2013PhRvD..88d4026H,2014ApJ...790...19K,2015PhRvD..91f4001T,2015PhRvD..91l4041D}.
The neutrino emissivity of the MNS-torus system is quite different from that of the black hole-torus system.
The MNS itself emits a large amount of neutrinos.
In addition, the torus matter does not accrete into the black hole, but stops at the surface of the MNS.
Such an inner region of the torus could contribute significantly to the neutrino emission.
Therefore, if the lifetime of MNSs is sufficiently long, a huge influence of neutrinos on the ejecta is expected \citep{2014MNRAS.441.3444M,2017arXiv170306216L}.
Recent numerical relativity simulations for the NS-NS mergers \citep{2015PhRvD..91f4059S,2016PhRvD..93l4046S} have indeed shown that the effect of neutrino irradiation is significant near the rotational axis.
However, their simulation time is not so long as the accretion time of the torus ($\gtrsim 100$ ms).
Hence, to explore the entire effects of neutrinos a longer-term simulation is required.

In \cite{2015ApJ...813...38R}, the pair-annihilation heating rate in a MNS-torus system was calculated with a Monte-Carlo method using snapshots of the pseudo-Newtonian hydrodynamics simulations \citep{2014MNRAS.441.3444M} as their background configurations.
\cite{2017arXiv170102017P} also calculated the heating rate with a ray-tracing method as a post process of the Newtonian hydrodynamics simulations for the NS-NS merger \citep{2014MNRAS.443.3134P}.
It was suggested that the total energy deposited by the pair-annihilation heating is not large enough to account for the typical energy of observed SGRB events, although the presence of the MNS does increase the pair-annihilation rate.
\cite{2017arXiv170102017P} concluded that much larger neutrino luminosity is required to explain most of the observed SGRBs.
Compared with the Newtonian simulation, however, the merger becomes more violent in simulations with general relativity because of the stronger gravitational effects.
Hence, the temperature of the merger remnant becomes higher than that with Newtonian gravity.
Therefore, their Newtonian simulation may underestimate the neutrino luminosity.
Moreover, \cite{2017arXiv170102017P} employed a stiff EOS, referred to as TM1 \citep{1998NuPhA.637..435S,1998PThPh.100.1013S, 2012ApJ...748...70H}.
For the stiff EOS, the remnant MNS is less compact, and hence, the neutrino luminosity is lower than that with softer EOSs.
The neutrino pair-annihilation heating could be more efficient in the general relativistic simulation with softer EOS.

Motivated by the above considerations, in this paper, we investigate neutrino-driven ejecta from the MNS-torus system.
We perform long-term fully general relativistic simulations with an approximate neutrino transport in axial symmetry using the initial condition based on a three-dimensional numerical relativity merger simulation.
We also investigate the effects of the neutrino-antineutrino pair-annihilation heating, which is not taken into account in our previous studies \citep{2015PhRvD..91f4059S,2016PhRvD..93l4046S}.
The outline of this paper is as follows.
A brief description of the simulation setting and the initial condition which we employ are given in Sec.~2.
Then, the results of the present simulations are shown in Sec.~3.
Finally, discussions and the conclusions are given in Secs.~4 and 5, respectively.
Throughout this paper, we use the units of $c=1=G$, where $c$ and $G$ are the speed of light and the gravitational constant, respectively.

\section{Method}

\subsection{Einstein's Equation}
In our simulation, Einstein's equation is solved in a Baumgarte-Shapiro-Shibata-Nakamura (BSSN)-puncture formalism \citep{1995PhRvD..52.5428S,1999PhRvD..59b4007B,2006PhRvL..96k1101C,2006PhRvL..96k1102B}.
In our formalism, we evolve the conformal three-metric $\tilde{\gamma}_{ij}=\gamma^{-1/3} \gamma_{ij}$, the conformal factor $W=\gamma^{-1/6}$ \citep{2008PhRvD..77f4010M}, the trace of the extrinsic curvature $K=\gamma^{ij}K_{ij}$, the trace-free part of the extrinsic curvature $\tilde{A}_{ij}=\gamma^{-1/3}(K_{ij} - \gamma_{ij}K/3)$, and the auxiliary variable $F_i=\delta^{jk}\del_j \tilde{\gamma}_{ki}$.
Here, $\gamma_{\alpha\beta}=g_{\alpha\beta}+n_\alpha n_\beta$ is the induced metric, where $g_{\alpha\beta}$ and $n_\alpha$ are the spacetime metric and the timelike unit vector normal to the time slice, respectively, $\gamma=\det(\gamma_{ij})$, and $\det(\tilde{\gamma}_{ij})$ is assumed to be unity.
We adopt the so-called cartoon method \citep{2001IJMPD..10..273A,2003PhRvD..67b4033S} to impose axially symmetric conditions for the geometric quantities.
We evaluate the spatial derivative by a fourth-order central finite differencing.
We add sixth order Kreiss-Oliger-type dissipation terms \citep{2008PhRvD..77b4027B} to the evolution equations of the geometrical variables for the stability of numerical evolution.
For the gauge condition, we adopt dynamical lapse \citep{2003PhRvD..67h4023A} and shift \citep{2003ApJ...595..992S} conditions as
\begin{align}
\del_t \alpha &=-2\alpha K,\\
\del_t \beta^i &= \frac{3}{4} \tilde{\gamma}^{ij}(F_j+\del_t F_j \Delta t),
\end{align}
where $\alpha$ and $\beta_i$ are the lapse function and the shift vector, respectively, and $\Delta t$ is the time-step interval.

\subsection{Neutrino-radiation Hydrodynamics Equations}

For exploring the neutrino-driven outflow from the NS-NS merger remnant, we need to solve hydrodynamics equations together with neutrino radiation transfer equations.
In this work, we divide neutrinos into ``streaming" and ``trapped" components.
For streaming-neutrinos, we employ the so-called M1-closure scheme based on Thorne's moment formalism~\citep{1981MNRAS.194..439T,2011PThPh.125.1255S}.
On the other hand, for trapped-neutrinos, we employ a leakage-based scheme \citep[for the leakage scheme, see, e.g.,][]{1988PhR...163...95C}.
In the following, we give a brief description of our formulation.
In this work, we consider three species of neutrinos: electron neutrinos $\nu_e$, electron antineutrinos $\bar{\nu}_e$, and the other neutrino species $\nu_x$, which represents all of the heavy-lepton neutrinos $\nu_\mu$, $\bar{\nu}_\mu$, $\nu_\tau$, and $\bar{\nu}_\tau$.

The basic equation of the fluid is derived from the energy-momentum conservation equation.
On the other hand, the basic equation of neutrinos is derived from the energy-integrated, first moment of Boltzmann's equation.
They are written, respectively, as
\begin{align}
\nabla _\beta T_{\rm (fluid)}^{\alpha\beta} &= -Q^\alpha = - \sum_i Q^\alpha_{\nu_i},\label{eq:fluid}\\
\nabla _\beta T_{(\nu_i)}^{\alpha\beta} &= Q^\alpha_{\nu_i}, \label{eq:rad}
\end{align}
where $T_{\rm (fluid)}^{\alpha\beta}$ and $T_{(\nu_i)}^{\alpha\beta}$ are the energy-momentum tensors of the fluid and the $i$-th species of neutrinos, respectively, and  $Q^\alpha_{\nu_i}$ is the energy-momentum source term of the $i$-th species of neutrinos, which is determined by weak-interaction processes.
Here, the total energy-momentum tensor is $T^{\alpha\beta}_{\rm (tot)} = T_{\rm (fluid)}^{\alpha\beta}+\sum_i T_{(\nu_i)}^{\alpha\beta}$, which satisfies $\nabla_\beta T^{\alpha\beta}_{\rm (tot)}=0$.

In our scheme, the energy-momentum tensor of neutrinos is phenomenologically decomposed as
\begin{align}
T_{(\nu_i)}^{\alpha\beta} = T_{(\nu_i,{\rm T})}^{\alpha\beta} + T_{(\nu_i,{\rm S})}^{\alpha\beta},
\end{align}
where $T_{(\nu_i,{\rm T})}^{\alpha\beta}$ and $T_{(\nu_i,{\rm S})}^{\alpha\beta}$ are the contributions of trapped- and streaming- neutrinos, respectively.
We assume that a part of neutrinos produced at a rate $Q^\alpha_{\nu_i}$ becomes streaming-neutrinos at a leakage rate $Q_{({\rm leak})\nu_i}^\alpha$ \citep[for the detail of this term, see][]{2012PTEP.2012aA304S}, and the other part becomes trapped-neutrinos at a rate $Q^\alpha_{\nu_i}-Q^\alpha_{({\rm leak})\nu_i}$.

We assume that trapped-neutrinos are tightly coupled with the fluid.
Then the energy-momentum tensor of trapped-neutrinos can be written by that of the fluid of relativistic Fermi particles, and thus, we decompose the total energy-momentum tensor as
\begin{align}
T^{\alpha\beta}_{\rm (tot)} = T^{\alpha\beta} + \sum_i T_{(\nu_i, {\rm S})}^{\alpha\beta},
\end{align}
where $T^{\alpha\beta} = T_{\rm (fluid)}^{\alpha\beta}+\sum_i T_{(\nu_i, {\rm T})}^{\alpha\beta}$ is the energy-momentum tensor composed of the sum of the fluid and trapped-neutrinos.
It is written in the form
\begin{align}
T_{\alpha\beta} = \rho h u_\alpha u_\beta + P g_{\alpha\beta}.
\end{align}
Here $\rho$ is the baryon rest-mass density, $u^\alpha$ is the four-velocity of the fluid, $h=1+\varepsilon + P/\rho$ is the specific enthalpy, $P$ is the pressure, and $\varepsilon$ is the specific internal energy.
$P$ and $\varepsilon$ contain the contributions from baryons, electrons, positrons, and trapped-neutrinos.

We evolve the fluid variables in cylindrical coordinates $(R,\varphi,z)$.
Hereafter, we regard the $y=0$ plane of Cartesian coordinates $(x,y,z)$ as a meridional ($\varphi=0$) plane of cylindrical coordinates $(R,\varphi,z)$.
The Euler and energy equations can be written, respectively, as
\begin{widetext}
\begin{align}
\del_t (\rho_* \hat{e}) +\frac{1}{x}\del_k[x (\rho_* \hat{e} v^k +W^{-3} P(v^k+\beta^k))]&=\alpha\biggl[ W^{-3} S_{kl}K^{kl}- \gamma^{kl} \rho_* \hat{u}_k \del_l \ln \alpha +W^{-3} Q_{\rm (leak)}^\alpha n_\alpha\biggr],\\
\del_t (\rho_* \hat{u}_i) + \frac{1}{x}\del_k[x(\rho_* \hat{u}_i v^k+\alpha W^{-3}P\delta^k{}_i)]&= \alpha \biggl[-\rho_* \hat{e} \del_i\ln \alpha + \frac{1}{\alpha}\rho_* \hat{u}_k \del_i \beta^k \notag\\
&\hspace{20mm}+ \frac{1}{2}W^{-3} S_{kl}\del_i \gamma^{kl}-W^{-3} Q_{\rm (leak)}^\alpha \gamma_{\alpha i}\biggr],
\end{align}
\end{widetext}
where $\rho_* \equiv W^{-3} w\rho$, $\hat{u}_i\equiv h u_i$, $v^i \equiv u^i/u^t$, $\hat{e} \equiv hw - P/(w\rho)$, $S_{ij} \equiv T_{\alpha\beta}\gamma^\alpha{}_i \gamma^\beta{}_j=\rho h u_i u_j + P\gamma_{ij}$, $w\equiv -n_\alpha u^\alpha = \sqrt{1+\gamma^{ij}u_iu_j}$ is the Lorentz factor of the fluid, and $Q^\alpha_{\rm (leak)}$ is the sum of leakage source terms of all neutrino species.

The number density of baryons, electrons, and trapped-neutrinos is evolved by
\begin{align}
\nabla_\alpha (\rho u^\alpha) &=0,\label{eq:baryon}\\
\nabla_\alpha (\rho Y_e u^\alpha) &= {\cal R}_e,\label{eq:ye}\\
\nabla_\alpha (\rho Y_{\nu_i} u^\alpha) &= {\cal R}_{\nu_i} - {\cal R}_{({\rm leak})\nu_i},\label{eq:trapnu}
\end{align}
where ${\cal R}_e$ and ${\cal R}_{\nu_i}$ denote the source terms for the evolution of the lepton number density due to the weak interactions.
Here we assumed that a part of neutrinos produced at a rate ${\cal R}_{\nu_i}$ becomes trapped-neutrinos at a rate ${\cal R}_{\nu_i}- {\cal R}_{({\rm leak})\nu_i}$, where ${\cal R}_{({\rm leak})\nu_i}$ is the emission rate of neutrinos of the flavor $i$ due to the diffusion \citep[for the details, see][]{2012PTEP.2012aA304S}.
Equations.~\eqref{eq:baryon}--\eqref{eq:trapnu} are written in the following forms, respectively;
\begin{align}
\del_t \rho_* + \frac{1}{x}\del_k(x \rho_* v^k)&=0,\\
\del_t (\rho_* Y_e) + \frac{1}{x}\del_k(x \rho_*Y_e v^k)&=\alpha W^{-3} {\cal R}_e,
\end{align}
\begin{align}
\del_t (\rho_* Y_{\nu_i}) + \frac{1}{x}\del_k(x \rho_*Y_{\nu_i} v^k)&=\alpha W^{-3}({\cal R}_{\nu_i} - {\cal R}_{({\rm leak})\nu_i}).
\end{align}

For streaming-neutrinos, we decompose their energy-momentum tensor as 
\begin{align}
T_{(\nu_i,{\rm S})}{}_{\alpha\beta} = En_\alpha n_\beta + F_\alpha n_\beta + F_\beta n_\alpha +P_{\alpha\beta},
\end{align}
where $E \equiv T_{({\rm S})}{}_{\mu\nu}n^\mu n^\nu$, $F_\alpha \equiv -T_{({\rm S})}{}_{\mu\nu}\gamma^\mu{}_\alpha n^\nu$, and $P_{\alpha\beta} \equiv T_{({\rm S})}{}_{\mu\nu}\gamma^\mu{}_\alpha \gamma^\nu{}_\beta$ are the energy density, energy flux, and spatial stress tensor of streaming-neutrinos, respectively.
$F_\alpha$ and $P_{\alpha\beta}$ satisfy $F_\alpha n^\alpha =0 = P_{\alpha\beta}n^\alpha$.
In the above, we omitted the subscripts of the species of neutrinos.
The evolution equations for $E$ and $F_i$ are written as
\begin{widetext}
\begin{align}
\del_t \tilde{E} + \frac{1}{x}\del_k [x(\alpha \tilde{F}^k-\beta^k \tilde{E})]=&\alpha \bigl[ \tilde{P}^{kl}K_{kl} -\tilde{F}^k\del_k \ln \alpha - W^{-3} Q_{\rm (leak)}^\alpha n_\alpha\bigr],\\
\del_t \tilde{F}_i + \frac{1}{x}\del_k [x(\alpha \tilde{P}_i{}^k-\beta^k \tilde{F}_i)]=&\alpha\biggl[-\tilde{E}\del_i \ln \alpha + \frac{1}{\alpha}\tilde{F}_k \del_i \beta^k +\frac{1}{2}\tilde{P}^{kl}\del_i\gamma_{kl} + W^{-3} Q_{\rm (leak)}^\alpha \gamma_{\alpha i}\biggr],
\end{align}
\end{widetext}
where $\tilde{E} \equiv W^{-3} E$, $\tilde{F}_i \equiv W^{-3} F_i$, and $\tilde{P}^{ij} \equiv W^{-3}P^{ij}$.
In this formalism, we have to impose a condition to determine $P^{ij}$.
For this, we use the so-called M1-closure relation \citep{1984JQSRT..31..149L,2007A&A...464..429G}.
In this closure relation, the spatial stress tensor is written as
\begin{align}
P^{ij}&=\frac{3\chi-1}{2}\left(P^{ij}\right)_{\rm thin}+\frac{3(1-\chi)}{2}\left(P^{ij}\right)_{\rm thick},
\end{align}
where $\left(P^{ij}\right)_{\rm thin}$ and $\left(P^{ij}\right)_{\rm thick}$ are the spatial stress tensor in the optically thin and thick limits, respectively, and are written as
\begin{align}
\left(P^{ij}\right)_{\rm thin} &= E \frac{F^i F^j}{\gamma_{kl}F^k F^l},\label{eq:pthin} \\
\left(P^{ij}\right)_{\rm thick}&= \frac{E}{2w^2+1}\left[(2w^2-1)\gamma^{ij} -4V^iV^j \right] \notag\\
&+\frac{1}{w} \left[F^i V^j +F^j V^i\right] \notag\\
&+\frac{2F^k u_k}{w(2w^2+1)}\left[ -w^2\gamma^{ij} +V^iV^j \right], \label{eq:pthick}
\end{align}
where $V^i = \gamma^{ij} u_j$ \citep{2011PThPh.125.1255S}.
$\chi$ is the so-called variable Eddington factor, which is a function of a normalized flux $f^2 = h_{\alpha\beta}H^\alpha H^\beta/J^2$ and we choose the following form:
\begin{align}
\chi = \frac{3+4f^2}{5+2\sqrt{4-3f^2}}.
\end{align}
Here we defined the energy density $J$, energy flux $H^\alpha$, and spatial stress-tensor $L^{\alpha\beta}$ of streaming-neutrinos in the fluid rest frame by $J=T_{({\rm S})}^{\alpha \beta}u_\alpha u_\beta$, $H^\alpha=-T_{({\rm S})}^{\gamma\beta}u_\beta h^\alpha{}_\gamma$, and $L^{\alpha\beta}=T_{({\rm S})}^{\gamma\delta}h^\alpha{}_\gamma h^\beta{}_\delta$, respectively, where $h_{\alpha\beta}=g_{\alpha\beta}+u_\alpha u_\beta$ is a projection operator.

\subsection{Microphysics}
\subsubsection{Equation of State}
In this paper, for the nuclear equation of state (EOS), we adopt a tabulated EOS referred to as DD2 \citep{2014ApJS..214...22B}, which is the same EOS as one of those used in three-dimensional simulations for the NS-NS merger \citep{2015PhRvD..91f4059S}.
The remnant of the NS-NS merger of the typical total mass ($\sim 2.7M_\odot$) is a long-lived MNS, and hence, we can explore the long-term phenomena associated with the MNS surrounded by a torus.

The original table of this EOS covers only ranges of density $10^{3.2}-10^{16.2}\ {\rm g\ cm^{-3}}$ and temperature $10^{-1}-10^{2.2}$ MeV, but the density and the temperature of the ejecta become lower than the lowest values of the table.
For treating such low density and temperature regions, we extend the table to ranges with $10^{-0.8}-10^{16.2}\ {\rm g\ cm^{-3}}$ and $10^{-3}-10^{2.2}$ MeV using an EOS by \cite{2000ApJS..126..501T}.
This EOS includes contributions of nucleons, heavy nuclei, electrons, positrons and photons to the pressure and the internal energy.

\subsubsection{Weak Interaction} \label{sec:weak}
The source terms of Eqs.~\eqref{eq:fluid} and \eqref{eq:rad} are
\begin{align}
Q_{\nu_e}^\alpha =& \biggl(Q_{{\rm brems},{\nu_e}}^{(-)} + Q_{{\rm pair},{\nu_e}}^{(-)}+ Q_{{\rm plasm},{\nu_e}}^{(-)}+Q_{\rm ec}^{(-)}\biggr)u^\alpha\notag\\
&-Q_{{\rm pair},{\nu_e}}^{(+)}{}^\alpha-Q_{{\rm abs},\nu_e}^{(+)}{}^\alpha,  \\
Q_{\bar{\nu}_e}^\alpha =& \biggl(Q_{{\rm brems},{\bar{\nu}_e}}^{(-)} + Q^{(-)}_{{\rm pair},{\bar{\nu}_e}}+ Q^{(-)}_{{\rm plasm},{\bar{\nu}_e}}+Q^{(-)}_{\rm pc} \biggr)u^\alpha\notag\\
&-Q_{{\rm pair},{\bar{\nu}_e}}^{(+)}{}^\alpha-Q_{{\rm abs},\bar{\nu}_e}^{(+)}{}^\alpha, \\
Q_{\nu_x}^\alpha =& \biggl(Q_{{\rm brems},{\nu_x}}^{(-)} + Q^{(-)}_{{\rm pair},{\nu_x}}+ Q^{(-)}_{{\rm plasm},{\nu_x}}\biggr)u^\alpha\notag\\
&-Q_{{\rm pair},{\nu_x}}^{(+)}{}^\alpha.
\end{align}
Here, $Q_{{\rm brems},{\nu_i}}^{(-)}$, $Q_{{\rm pair},{\nu_i}}^{(-)}$, $Q_{{\rm plasm},{\nu_i}}^{(-)}$, $Q_{\rm ec}^{(-)} $, and $Q_{\rm pc}^{(-)}$ are the neutrino cooling (emission) rates due to the nucleon-nucleon Bremsstrahlung, the electron-positron pair-annihilation, the plasmon decay, the electron-capture, and the positron-capture processes, respectively.
$Q_{{\rm pair},{\nu_i}}^{(+)}{}^\alpha$, $Q_{{\rm abs},\nu_e}^{(+)}{}^\alpha$, and $Q_{{\rm abs},\bar{\nu}_e}^{(+)}{}^\alpha$ are the matter-heating source terms due to the neutrino-antineutrino pair-annihilation, the electron-neutrino and electron-antineutrino absorption processes, respectively.
For electron and positron capture processes, we use the rate in \cite{1985ApJ...293....1F}.
For pair-production processes, we use the rates in \cite{1986ApJ...309..653C} for electron-positron pair-annihilation, \cite{1996A&A...311..532R} for plasmon-decay, and \cite{2006NuPhA.777..356B} for nucleon-nucleon Bremsstrahlung, respectively.
Except for the neutrino pair-annihilation process, the rates adopted in the present simulations are the same as those adopted in the simulations for three-dimensional NS-NS mergers \citep{2015PhRvD..91f4059S}.
The explicit forms of the cooling rates ($Q_{{\rm brems},{\nu_i}}^{(-)}$, $Q_{{\rm pair},{\nu_i}}^{(-)}$, $Q_{{\rm plasm},{\nu_i}}^{(-)}$, $Q_{\rm pc}^{(-)} $, and $Q_{\rm ec}^{(-)}$) are found in \cite{2012PTEP.2012aA304S}.

In our moment formalism, the source term due to the neutrino pair-annihilation process is written (in $\hbar=1$ unit) as
\begin{widetext}
\begin{align}
Q^{(+)}_{{\rm pair},\nu_i}{}^\alpha=\ & \biggl[\frac{C_{\nu_i \bar{\nu}_i}^{\rm pair} G_{\rm F}^2}{3\pi} \left<\omega\right>_{\rm pair} \bigl(J \bar{J}-2H^\beta \bar{H}_\beta+L^{\beta\gamma} \bar{L}_{\beta\gamma}\bigr) e^{-\tau_{\nu_i}}\biggr] u^\alpha \notag\\
=\ & Q^{(+)}_{{\rm pair},\nu_i} u^\alpha,\label{eq:pairrate}
\end{align}
\end{widetext}
where $\left<\omega\right>_{\rm pair}$ is the average energy of streaming-neutrinos that annihilate, and the quantities with bar imply those of antineutrinos.
The exponential factor in the first line of Eq.~\eqref{eq:pairrate} is introduced to suppress the heating rate in the optically thick region because the pair-annihilation heating balances with electron-positron pair-annihilation cooling in such a region.
$G_{\rm F}$ is the Fermi coupling constant and $C_{\nu_i \bar{\nu}_i}^{\rm pair}$ is written by the Weinberg angle $\theta_{\rm W}$ as $C_{\nu_i \bar{\nu}_i}^{\rm pair} = 1\pm 4\sin^2\theta_{\rm W} + 8\sin^4\theta_{\rm W}$, where the plus and minus signs denote neutrinos of electron and heavy lepton types, respectively.
We use the value $\sin^2\theta_{\rm W}\approx 0.2319$.
The detailed derivation of Eq.~\eqref{eq:pairrate} is described in Appendix~\ref{app:pair}.

Because we assume an analytic closure relation, the evaluation of the pair-annihilation rate in our simulation could be significantly different from the true rate.
In this paper, the uncertainty in the pair-annihilation heating rate is explored by performing a simulation in which the pair-annihilation heating rate is calculated assuming the isotropic momentum-space angular distribution for neutrinos.
Under this assumption, the pair-annihilation source term $Q^{(+)}_{{\rm pair/Iso},\nu_i}\ (i=e,x)$ is written as
\begin{align}
Q^{(+)}_{{\rm pair/Iso},\nu_i}= & \frac{C_{\nu_i \bar{\nu}_i}^{\rm pair} G_{\rm F}^2}{3\pi} \left<\omega\right>_{\rm pair} \frac{4}{3}J \bar{J} e^{-\tau_{\nu_i}}.\label{eq:pairiso}
\end{align}
In this model, the pair-annihilation heating rate may be incorporated in an optimistic manner (see Sec.~\ref{subsec:vol-int} for discussion).

For the source terms of Eqs.~\eqref{eq:ye} and \eqref{eq:trapnu}, we employ the following equations:
\begin{align}
{\cal R}_e &= {\cal R}_{{\rm abs},\nu_e}-{\cal R}_{{\rm abs},\bar{\nu}_e}+{\cal R}_{\rm pc}-{\cal R}_{\rm ec},\\
{\cal R}_{\nu_e} &= {\cal R}_{{\rm brems},{\nu_e}} + {\cal R}_{{\rm pair},{\nu_e}}+ {\cal R}_{{\rm plasm},{\nu_e}}+{\cal R}_{\rm ec}, \\
{\cal R}_{\bar{\nu}_e} &= {\cal R}_{{\rm brems},{\bar{\nu}_e}} + {\cal R}_{{\rm pair},{\bar{\nu}_e}}+ {\cal R}_{{\rm plasm},{\bar{\nu}_e}}+{\cal R}_{\rm pc}, \\
{\cal R}_{\nu_x} &= {\cal R}_{{\rm brems},{\nu_x}} + {\cal R}_{{\rm pair},{\nu_x}}+ {\cal R}_{{\rm plasm},{\nu_x}}.
\end{align}
For these rates except for the electron (anti)neutrino absorption processes, we employ the rates in \cite{1985ApJ...293....1F}, \cite{1986ApJ...309..653C}, \cite{1996A&A...311..532R}, and \cite{2006NuPhA.777..356B}, which are the same as in \cite{2012PTEP.2012aA304S}.

The source terms due to the electron (anti)neutrino absorption processes are written as
\begin{align}
Q_{{\rm abs},\nu_e}^{(+)}{}^\alpha &\approx \frac{(1+3g_{\rm A}^2)G_{\rm F}^2}{\pi} \frac{\rho X_n}{m_u} \left<\omega_{\nu_e} \right>_{\rm abs}^2 (Ju^\alpha + H^\alpha) e^{-\tau_{\nu_e}},\label{eq:Qabsn}\\
Q_{{\rm abs},\bar{\nu}_e}^{(+)}{}^\alpha &\approx \frac{(1+3g_{\rm A}^2)G_{\rm F}^2}{\pi} \frac{\rho X_p}{m_u} \left<\omega_{\bar{\nu}_e} \right>_{\rm abs}^2 (\bar{J}u^\alpha + \bar{H}^\alpha)e^{-\tau_{\bar{\nu}_e}},\\
{\cal R}_{{\rm abs},{\nu_e}} &\approx \frac{(1+3g_{\rm A}^2)G_{\rm F}^2}{\pi}\frac{\rho X_n}{m_u} \left<\omega_{{\nu_e}} \right>_{\rm abs} J e^{-\tau_{\nu_e}},\\
{\cal R}_{{\rm abs},\bar{\nu}_e} &\approx \frac{(1+3g_{\rm A}^2)G_{\rm F}^2}{\pi}\frac{\rho X_p}{m_u} \left<\omega_{\bar{\nu}_e} \right>_{\rm abs} \bar{J}e^{-\tau_{\bar{\nu}_e}},\label{eq:Rabsa}
\end{align}
where $\left<\omega \right>_{\rm abs}$ is the average energy of streaming-neutrinos, $m_u$ is the atomic mass unit, $g_{\rm A}\approx 1.26$ is the axial vector coupling strength, and $X_n$ and $X_p$ are the mass fractions of neutrons and protons, respectively.
For the same reason as in the pair-annihilation heating, the exponential factors in Eqs.~\eqref{eq:Qabsn}--\eqref{eq:Rabsa} are introduced to suppress these rates in the optically thick region, in which these rates balance with those of their inverse processes.
The detailed derivation for these rates is described in Appendix~\ref{app:abs}.

\subsubsection{Average Energy Estimation}
For evaluating the average energy of streaming-neutrinos $\left<\omega\right>_{\rm pair}$ in our energy-integrated radiation transfer scheme, we assume that streaming-neutrinos have the Fermi-Dirac-type energy distribution, i.e.,
\begin{align}
f_{\nu}(\omega) = \frac{1}{e^{\omega/T_\nu-\eta_\nu}+1}.
\end{align}
Then the energy density of neutrinos in the fluid rest frame $J$, the ``temperature" of the neutrino $T_\nu$, and the normalized ``chemical potential" of streaming-neutrinos $\eta_\nu$ satisfy the relation
\begin{align}
J=\int d\omega \frac{4\pi \omega^3}{(2\pi)^3} \frac{1}{e^{\omega/T_\nu-\eta_\nu}+1} = \frac{T_\nu^4}{2\pi^2}F_3(\eta_\nu), \label{eq:Jetanu}
\end{align}
where $F_k(\eta)$ is the relativistic Fermi integral of order $k$ defined by
\begin{align}
F_k(\eta)=\int dx \frac{x^k}{e^{x-\eta}+1}.
\end{align}
We assumed that the ``temperature" of streaming-neutrinos $T_\nu$ is the same as the local temperature of the matter $T$.
Then we obtain the normalized ``chemical potential" of neutrinos, $\eta_\nu$, by solving Eq.~\eqref{eq:Jetanu}.
The average energy of streaming-neutrinos is then evaluated as
\begin{align}
\left<\omega\right>_{\rm pair} = \frac{F_4(\eta_\nu)}{F_3(\eta_\nu)}T. \label{eq:ave}
\end{align}
This average energy would be smaller than the real one because the temperature of neutrinos is in reality comparable to the temperature of their emission region, which is usually higher than that of the free-streaming region of neutrinos.
For this reason, we should keep in mind that the pair-annihilation heating rate would be evaluated by our scheme in a conservative manner.

In the same way as Eq.~\eqref{eq:ave}, we evaluate $\left<\omega \right>_{\rm abs}$ as
\begin{align}
\left<\omega\right>_{\rm abs} = \frac{F_3(\eta_\nu)}{F_2(\eta_\nu)}T, \label{eq:ave32}
\end{align}
where we simply derive the average energy of neutrinos assuming the Fermi-Dirac-type energy distribution.

We have to keep in mind that there are several uncertainties in the average energy estimation in our energy-integrated neutrino transport scheme.
In particular, the neutrino absorption reactions have strong energy dependence, so that the change of the average energy would affect the dynamics of the neutrino-driven ejecta.
For checking such uncertainties, we perform a simulation in which the average energy of neutrinos are estimated by
\begin{align}
\left<\omega\right>_{\rm abs} = \sqrt{\frac{F_5(\eta_\nu)}{F_3(\eta_\nu)}} T \label{eq:ave53}
\end{align}
instead of Eq.~\eqref{eq:ave32}.
This is derived when we take the energy dependence of the absorption reaction into account, and factor out the neutrino energy density of Eq.~\eqref{eq:Jetanu}.
We give a detail for deriving Eq.~\eqref{eq:ave53} in Appendix~\ref{app:abs}.

\subsection{Initial Condition} \label{subsec:initial}
We mapped three-dimensional simulation data for a NS-NS merger \citep{2015PhRvD..91f4059S} into two-dimensional data as the initial condition for the axisymmetric simulation.
Although the merger remnant has a non-axisymmetric structure in an early phase after the merger, it gradually relaxes to a nearly axisymmetric structure.
It also approaches a quasi-stationary state.
In this work, we employ the remnant at about 50 ms after the onset of the merger as an initial condition.
In such a phase, the merger remnant is approximately in an axisymmetric quasi-stationary state.

We generate axisymmetric data from three-dimensional data by taking average for $\rho_*$,  $\hat{e}$, $\hat{u}_i$, $Y_e$, and $Y_{\nu_i}$ by
\begin{align}
\rho_* &= \int \frac{d\varphi}{2\pi} \rho_*^{\rm (3D)},\\
\hat{e}&=\frac{1}{\rho_*} \int \frac{d\varphi}{2\pi} \rho_*^{\rm (3D)}\hat{e}^{\rm (3D)},\\
\hat{u}_x &= \frac{1}{\rho_*}\int \frac{d\varphi}{2\pi} \rho_*^{\rm (3D)} (+\hat{u}_x^{\rm (3D)}\cos\varphi + \hat{u}_y^{\rm (3D)}\sin\varphi),\\
\hat{u}_\varphi &= \frac{1}{\rho_*}\int \frac{d\varphi}{2\pi} \rho_*^{\rm (3D)} (-\hat{u}_x^{\rm (3D)}\sin\varphi + \hat{u}_y^{\rm (3D)}\cos\varphi),\\
\hat{u}_z &= \frac{1}{\rho_*}\int \frac{d\varphi}{2\pi} \rho_*^{\rm (3D)} \hat{u}_z^{\rm (3D)},\\
Y_i &= \frac{1}{\rho_*}\int \frac{d\varphi}{2\pi} \rho_*^{\rm (3D)} Y_i^{\rm (3D)},
\end{align}
where the quantities with superscript (3D) imply those taken from the corresponding three-dimensional data.
After mapping, we solve the Hamiltonian and momentum constraint equations assuming the conformal flatness.
Employing the conformal flatness is reasonable because the values of $\tilde{\gamma}_{ij}-\delta_{ij}$ are fairly small at $\sim$50 ms after the merger; specifically the absolute magnitude of all the components is smaller than 0.02.

\begin{figure}[t]
\includegraphics[bb=0 0 360 360,width=\hsize]{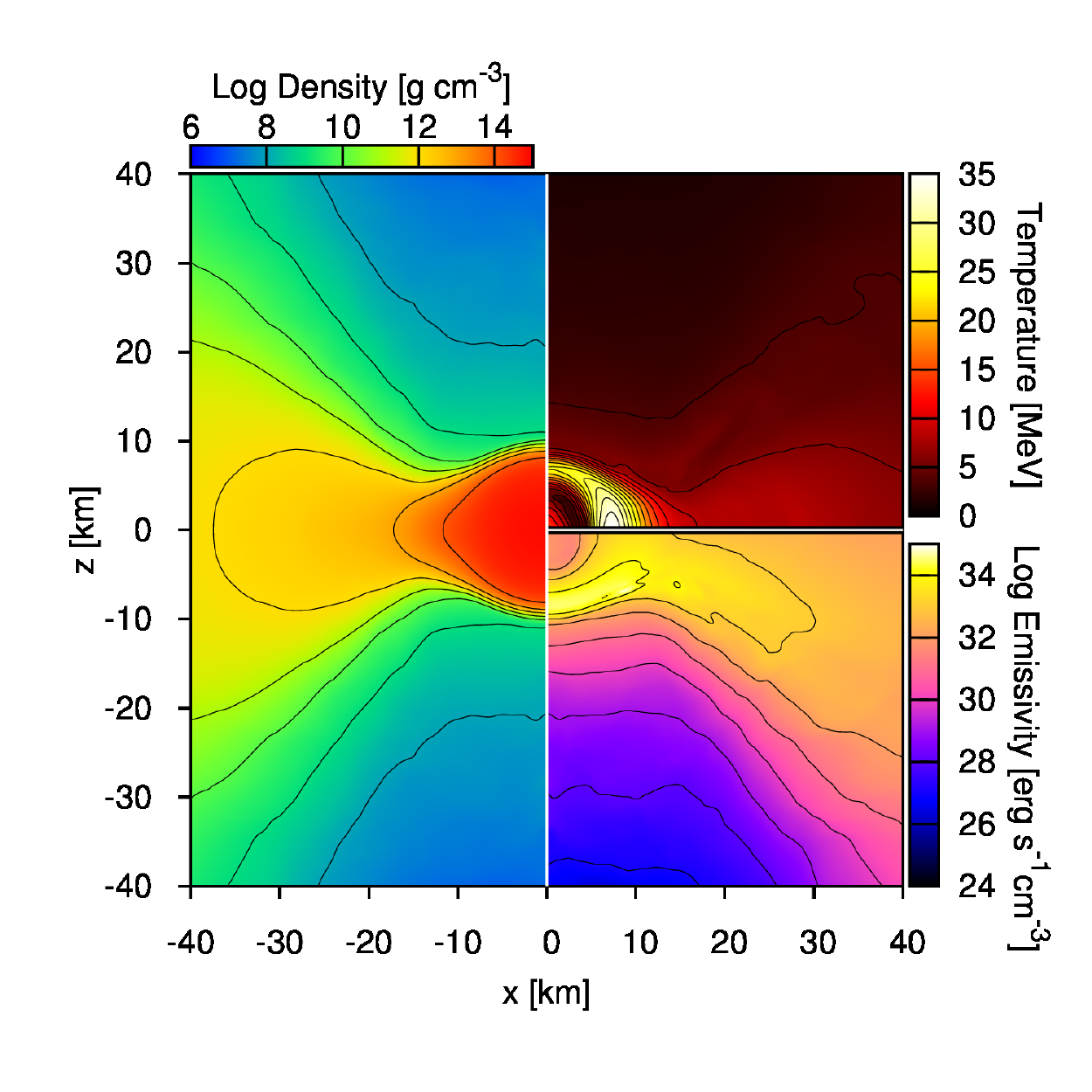}
\caption{
Initial thermodynamical configuration.
The rest-mass density (the second and the third quadrant), the temperature (the first quadrant), the flavor-summed effective neutrino emissivity (the fourth quadrant) are shown in the same $x$-$z$ plane.
The black curves denote contours of each quantity.
For the temperature, contours are linearly spaced with the interval of 2.5 MeV.
For the neutrino emissivity and the rest-mass density, they are logarithmically spaced with the intervals of 1.0 dex.
}
\label{fig:init}
\end{figure}

The configurations of initial thermodynamical quantities are shown in Fig.~\ref{fig:init}.
The MNS is surrounded by a massive torus of mass $\sim0.2M_\odot$.
Here we defined the mass of the torus as the baryonic mass in the region where the density is lower than $10^{13}\ {\rm g\ cm^{-3}}$.
The maximum temperature is $\sim$30 MeV near the equatorial surface region of the MNS.
The temperature of the inner region of the torus is $\sim5-7$ MeV.
The effective neutrino emissivity is highest on the polar surface of the MNS and in the inner region of the torus.

We summarize the models for our simulations in Table~\ref{tab:models}.
We refer to the fiducial model as DD2-135135-On-H.
For understanding the effects of the pair-annihilation heating, we also perform a simulation without the neutrino pair-annihilation (the model DD2-135135-Off-H).
In addition to these models, we perform simulations of the models DD2-135135-Iso-H and DD2-135135-On-H-53 to check the uncertainty of the heating rates due to the neutrino pair-annihilation and the neutrino absorption processes.
In the former model, the pair-annihilation heating rate is calculated by Eq.~\eqref{eq:pairiso} assuming the isotropic momentum-space angular distribution for neutrinos.
The latter is the simulation in which the average energy used in the neutrino absorption heating process is estimated by Eq.~\eqref{eq:ave53} instead of Eq.~\eqref{eq:ave32}.

\begin{table*}[t]
\caption{
List of the models. $M_{\rm NS}$ is the gravitational mass of each NS in the merger simulation of \cite{2015PhRvD..91f4059S}.
For $\Delta x_0$, $\delta$, $R_{\rm star}$, $N$, and $L$, see the text in Sec.~\ref{sec:grid}.
}
\begin{center}
\begin{tabular*}{\hsize}{@{\extracolsep{\fill}}lccccccccc}
\hline \hline
Model & EOS &$M_{\rm NS}$& Pair-annihilation & $\nu$-Absorption & $\Delta x_0$& $\delta$ & $R_{\rm star}$& $N$ & $L$\\
&&$(M_\odot)$  & heating & & (m)& & (km)& & (km) \\
\hline
DD2-135135-On-H (fiducial,high)    &DD2& 1.35  & on & Eq.\eqref{eq:ave32} & 150 & 0.0075 & 30 & 951 & 5440\\
DD2-135135-Off-H &DD2& 1.35  & off & Eq.\eqref{eq:ave32} & 150 & 0.0075 & 30  & 951 & 5440\\
DD2-135135-Iso-H &DD2& 1.35  & on, isotropic & Eq.\eqref{eq:ave32} & 150 & 0.0075 & 30 & 951 & 5440\\
DD2-135135-On-H-53 &DD2& 1.35  & on & Eq.\eqref{eq:ave53} & 150 & 0.0075 & 30 & 951 & 5440\\
\hline
DD2-135135-On-M (medium)&DD2& 1.35  & on & Eq.\eqref{eq:ave32} & 200 & 0.0075 & 30 & 871 & 5790 \\
DD2-135135-On-L (low) &DD2& 1.35  & on & Eq.\eqref{eq:ave32} & 250 & 0.0075 & 30 & 809 & 5690 \\
\hline
\end{tabular*}
\end{center}
\label{tab:models}
\end{table*}

\subsection{Grid Setting} \label{sec:grid}
We adopt a non-uniform grid in which the grid spacing is increased according to the rule
\begin{align}
\Delta x_{j+1} &= 
\begin{cases}
\Delta x_j & (x_j\le R_{\rm star}) \\
(1+\delta)\Delta x_j & (x_j > R_{\rm star}),
\end{cases}\notag\\
\Delta z_{l+1} &= 
\begin{cases}
\Delta z_l & (z_l\le R_{\rm star}) \\
(1+\delta)\Delta z_l & (z_l > R_{\rm star}),
\end{cases}
\end{align}
where $\Delta x_j \equiv x_{j+1} - x_j$, $\Delta z_l \equiv z_{l+1} - z_l$, $\delta$ and $R_{\rm star}$ are constants, and $0\le j \le N$ ($0\le l \le N$).
Here, $N+1$ is the total grid number for one direction.
In this grid, a uniform grid with the grid spacing $\Delta x_0$ is adopted in the inner region $0\le x \le R_{\rm star}$ and $0\le z \le R_{\rm star}$ to resolve the MNS.
The convergence of numerical results is examined by performing simulations with coarser grid resolutions (DD2-135135-On-M and DD2-135135-On-L, see Table~\ref{fig:init}).
The values of the innermost grid spacing $\Delta x_0$, the constant $\delta$, the size of the region in which the uniform grid is adopted $R_{\rm star}$, the grid number $N$, and the size of computational domain $L$ are also tabulated in Table~\ref{tab:models}.

\section{Result}
\subsection{Dynamics of the System}
\begin{figure*}[t]
\begin{center}
\includegraphics[bb=0 0 360 540, width=0.45\hsize]{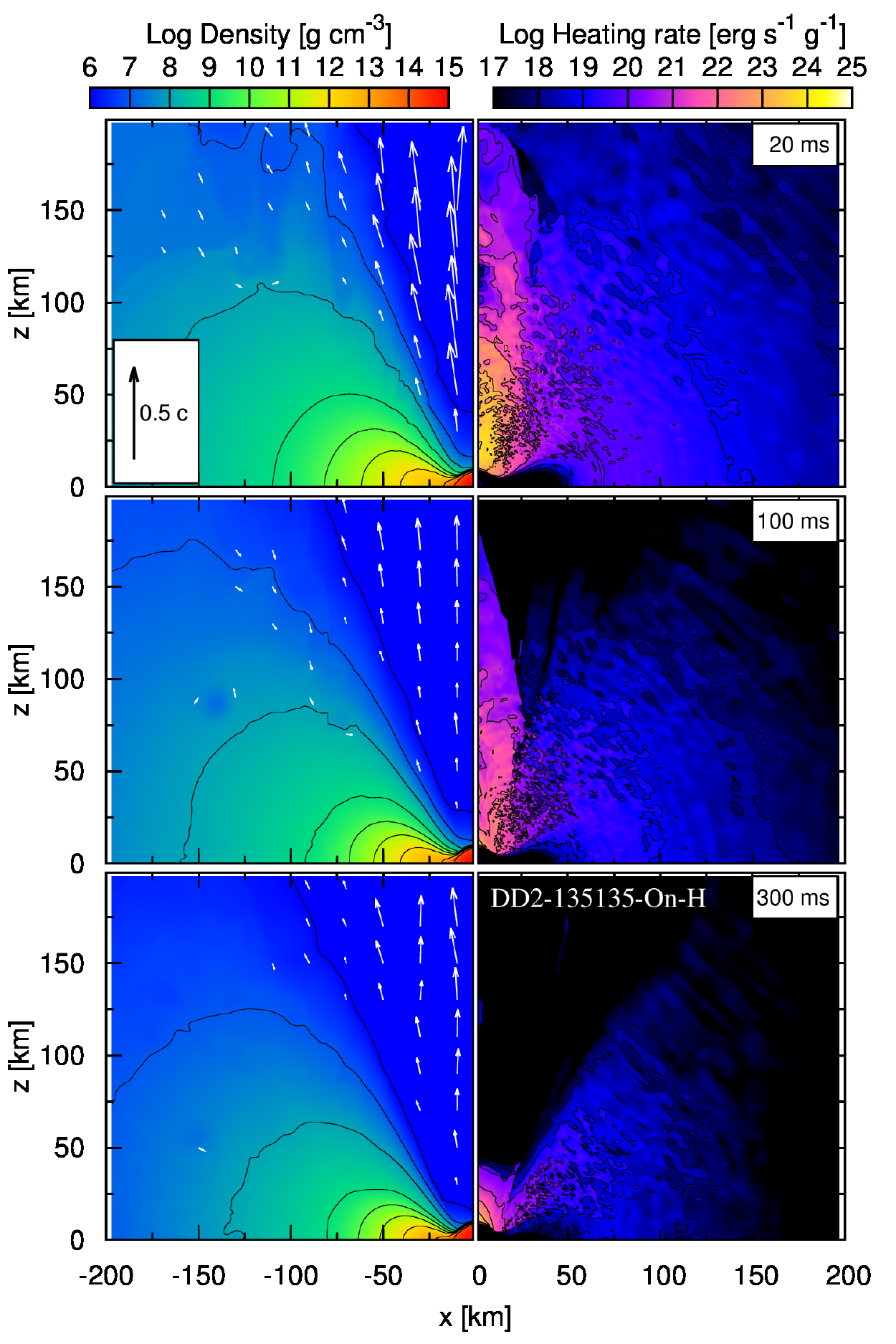}
\includegraphics[bb=0 0 360 540,width=0.45\hsize]{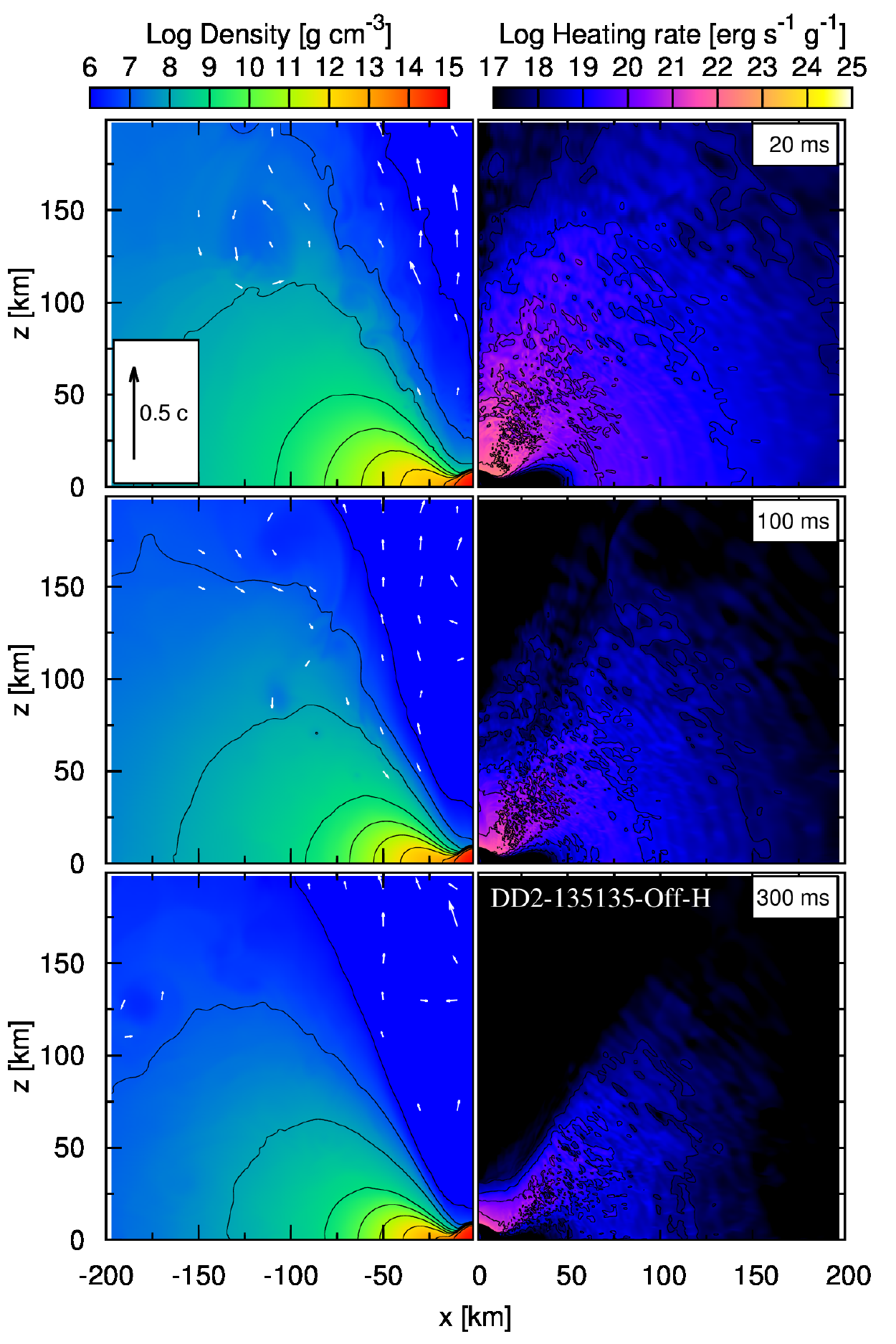}
\caption{
Left: Snapshots of the rest-mass density, the poloidal velocity fields $(v^x,v^z)$, and the specific heating rate for the model DD2-135135-On-H at $t=$ 20 ms (top), 100 ms (middle), and 300 ms (bottom) in the meridional plane.
The velocity field is plotted only for the case that the velocity is larger than 0.03 $c$.
Right: Same as the left panel, but for the model DD2-135135-Off-H.
For the figures of the rest-mass density, the black curves denote logarithmically spaced contours with the intervals of 1.0 dex.
}
\label{fig:den}
\end{center}
\end{figure*}

First, we briefly summarize the dynamics of the fiducial model DD2-135135-On-H.
For an early phase of evolution ($t\lesssim 50$ ms, where $t$ is the time after mapping from the 3D simulation), a strong outflow is launched in the vicinity of the rotational axis.
This is driven primarily by the neutrino pair-annihilation heating in the dilute matter.
The neutrino absorption heating is subdominant in the fiducial model.
The evolution of the rest-mass density, the velocity vector, and the specific heating rate is displayed in the left panel of Fig.~\ref{fig:den}.
This clearly shows that the outflow is launched around the rotational axis.
The maximum velocity of the ejecta is $\sim 0.5\ c$.
We also display snapshots of the rest-mass density, the velocity vector, and the specific heating rate in the absence of the pair-annihilation heating (DD2-135135-Off-H) in the right panel of Fig.~\ref{fig:den}.
For this case, an outflow is launched only by the neutrino absorption heating, and the ejecta velocity is $0.1-0.2\ c$.
As found in the right halves of the both panels in Fig.~\ref{fig:den}, the specific heating rate is considerably different between the two simulations.
The specific heating rate in the polar region of the model DD2-135135-On-H is much higher than that of the model DD2-135135-Off-H because the pair-annihilation heating dominates the other heating processes (i.e., the absorption of electron neutrinos and electron antineutrinos on nucleons).
This illustrates that the high-speed ejecta is launched by the strong pair-annihilation heating.

At $t\sim$ 50 ms, the maximum velocity of the ejecta goes down to $\sim 0.2\ c$ even in the presence of the pair-annihilation heating.
This is because the neutrino pair-annihilation heating rate decreases with time as found in the middle panel of Fig.~\ref{fig:den} (and see also Fig.~\ref{fig:lumpair}).
The decrease of the pair-annihilation heating rate is caused by the decrease of the neutrino luminosity.

\subsection{Neutrino Luminosity, Pair-annihilation Rate, and Efficiency}\label{subsec:neutrino}
\begin{figure}[htbp]
\includegraphics[bb=0 0 360 504,width=\hsize]{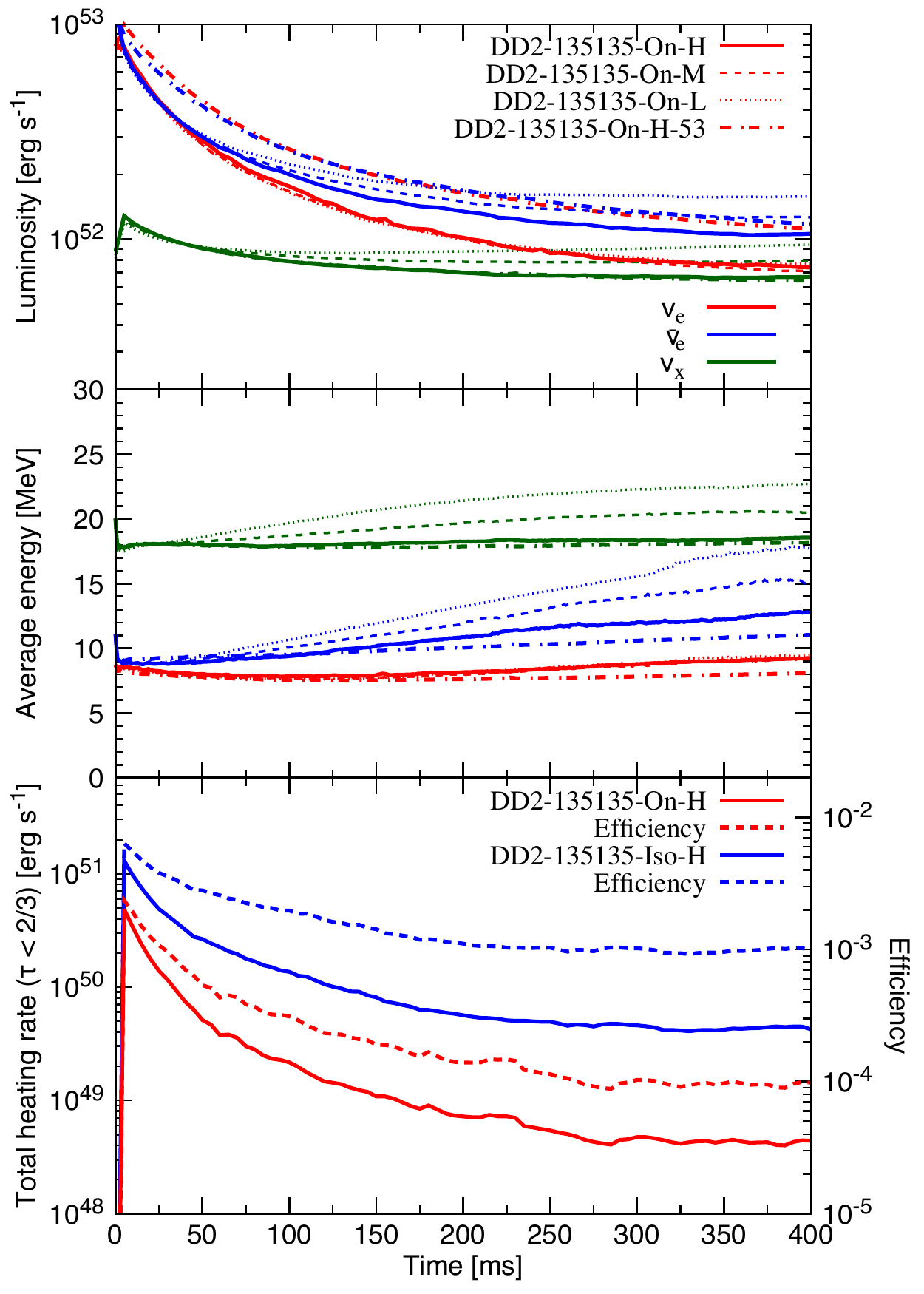}
\caption{
Top: Luminosity curves for electron neutrinos (red), electron antineutrinos (blue), and other neutrinos (green), respectively.
The thick solid, thin dashed, and thin dotted curves denote the results for three different resolution models DD2-135135-On-H (high-resolution), DD2-135135-On-M (medium-resolution) and DD2-135135-On-L (low-resolution), respectively.
The dot-dashed curves denote the result of the model DD2-135135-On-H-53, in which the average energy of neutrinos is estimated by Eq.~\eqref{eq:ave53} instead of Eq.~\eqref{eq:ave32}.
Middle: Average energies of individual neutrino species.
The line types and colors correspond to the resolutions and the neutrino species in the same way as in the top panel.
Bottom: Total heating rates due to neutrino pair-annihilation (solid curves) and the heating efficiencies (dashed curves) for the models DD2-135135-On-H (red) and DD2-135135-Iso-H (blue), respectively.
The definition for the heating rate is found in Eq.~\eqref{eq:qpairtot}.
}
\label{fig:lumpair}
\end{figure}

\begin{figure}[t]
\includegraphics[bb=0 0 360 540,width=\hsize]{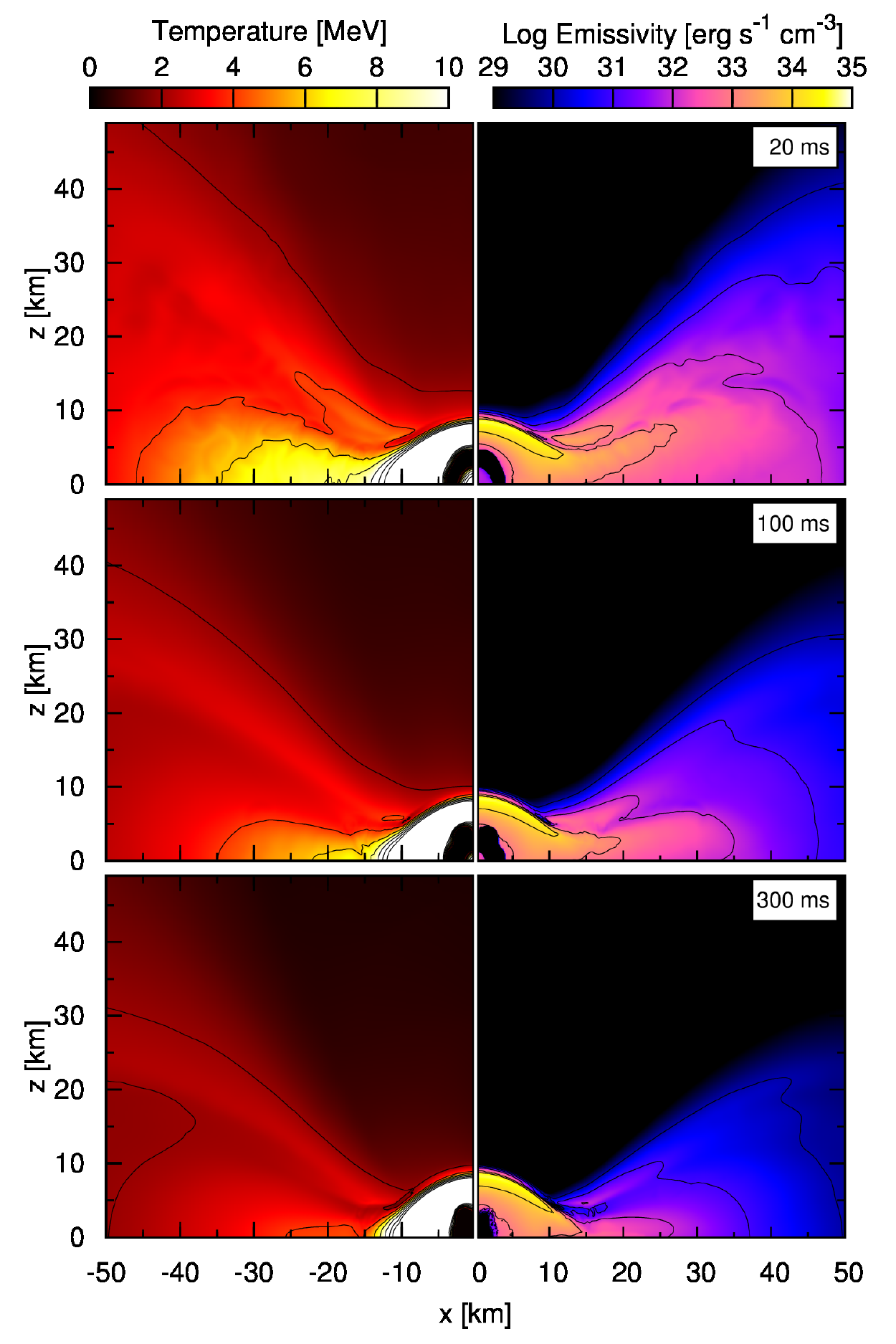}
\caption{
Snapshots of the temperature (left) and the neutrino emission cooling rate (right) for the model DD2-135135-On-H at $t=$ 20 ms (top), 100 ms (middle), and 300 ms (bottom) in the meridional plane.
The white region in the left panel denotes the region in which the temperature is higher than 10 MeV.
The black curves denote contours of each quantity.
For the temperature the spacing of the contour is 2.5 MeV.
For the cooling rate, the contour is logarithmically spaced with the intervals of 1.0 dex.
}
\label{fig:tem_cool}
\end{figure}

The top panel of Fig.~\ref{fig:lumpair} shows the luminosity curves of individual neutrinos as functions of time.
At the beginning of the simulation, the luminosity of electron-type neutrinos is $\sim 10^{53}\ {\rm erg\ s^{-1}}$, and an order of magnitude larger than those by proto-neutron stars formed during the typical supernova explosions \citep{2010A&A...517A..80F}.
In 100 ms, the luminosity decreases rapidly, because the temperature of the torus, which is a strong emitter of electron-type neutrinos, decreases from $\sim 8$ MeV to $\sim 5$ MeV (see Fig. \ref{fig:tem_cool}).
For $t \gtrsim 300$ ms, the luminosity settles into nearly constant in time as $\sim 10^{52}\ {\rm erg\ s^{-1}}$.
In this phase, the primary source of the neutrino emission is the MNS and the neutrino emissivity of the torus is much smaller than that of the MNS.

We define the total pair-annihilation heating rate outside the neutrinosphere by
\begin{align}
Q_{\rm pair,tot}&=\sum_i \int_{\tau_{\nu_i}<2/3} d\Sigma_\alpha Q^{(+)}_{{\rm pair},\nu_i}{}^\alpha \notag\\
&= \sum_i \int_{\tau_{\nu_i}<2/3} dxdz \ 2\pi x W^{-3} w Q^{(+)}_{{\rm pair},\nu_i}, \label{eq:qpairtot}
\end{align}
where $d\Sigma_\alpha = d^3 x \sqrt{\gamma} n_\alpha$ is the three-dimensional volume element on spatial hyper-surfaces.
The bottom panel of Fig.~\ref{fig:lumpair} shows $Q_{\rm pair,tot}$ (solid curve) as a function of time.
In the fiducial model, $Q_{\rm pair,tot}$ is $\sim10^{50}\ {\rm erg\ s^{-1}}$ at $t \sim 50$ ms, and it decreases with time as the neutrino luminosity decreases.
For $t \agt 300$ ms, it settles down to $\sim 10^{49}\ {\rm erg\ s^{-1}}$.

In the same panel, we also plot the efficiency of the pair-annihilation heating (dashed curves)
\begin{align}
\eta_{\rm pair} \equiv \frac{Q_{\rm pair,tot}}{L_{\nu,{\rm tot}}}
\end{align}
as a function of time \citep{1997A&A...319..122R,2004MNRAS.352..753S}.
Since $Q_{\rm pair,tot} \propto L_\nu^2$, the efficiency is proportional to the neutrino luminosity.
Therefore, it decreases with time and eventually settles to $\sim 10^{-4}$ in the quasi-stationary phase.

This efficiency is consistent with those obtained with a Monte-Carlo method in the central NS case \citep{2015ApJ...813...38R} and obtained with a multi-energy M1-scheme in the central black hole case \citep{2016ApJ...816L..30J} when the individual neutrino luminosity is $\sim 10^{52}\ {\rm erg\ s^{-1}}$.

\subsection{Properties of Neutrino-driven Ejecta}
We define the baryonic mass, the total energy, and the internal energy of the ejecta by
\begin{align}
M_{\rm b,ej} &= \int_{|u_t|>1} d\Sigma_\alpha \rho u^\alpha = \int_{|u_t|>1} dx dz\ 2\pi x \rho_*,\label{eq:mej}\\
E_{\rm tot,ej} &=\int_{|u_t|>1} d\Sigma_\alpha \rho \hat{e} u^\alpha= \int_{|u_t|>1} dx dz\ 2\pi x \rho_* \hat{e},\label{eq:etotej}\\
E_{\rm int,ej} &= \int_{|u_t|>1} d\Sigma_\alpha \rho \varepsilon u^\alpha = \int_{|u_t|>1} dx dz\ 2\pi x \rho_* \varepsilon,\label{eq:eintej}
\end{align}
where we supposed that fluid elements with $|u_t|>1$ are gravitationally unbound.
Using these quantities, we define the kinetic energy of the ejecta by
\begin{align}
E_{\rm kin,ej} &= E_{\rm tot,ej} - M_{\rm b,ej} - E_{\rm int,ej}.
\end{align}
We evaluate these quantities in the cylindrical region of $x \leq 2000$ km and $-2000$ km $\leq z \leq  2000$ km.
We also take into account the mass and energy gone outside this inner region by integrating the fluxes at the boundary of this region and adding them to Eqs.~\eqref{eq:mej}--\eqref{eq:eintej}.

\begin{figure}[t]
\begin{center}
\includegraphics[bb=0 0 360 360,width=\hsize]{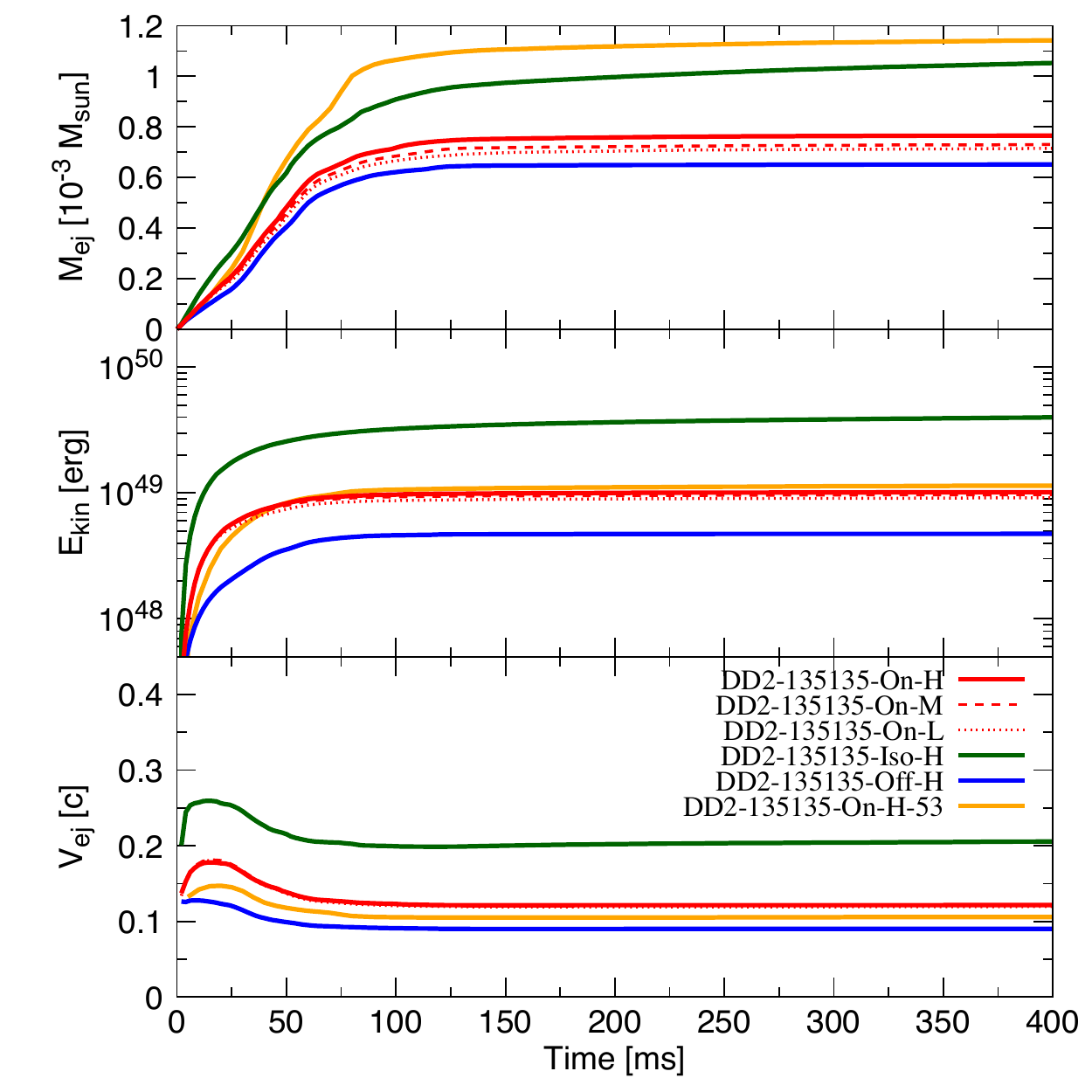}
\caption{
The total mass (top), the total kinetic energy (middle), and the average velocity (bottom) of the ejecta, respectively.
The solid red thick curves denote the results for the fiducial model DD2-135135-On-H, and the blue curves denote the result for the model DD2-135135-Off-H,  in which the heating process is not taken into account.
The thin dashed and thin dotted curves denote the results for the low-resolution models DD2-135135-On-M and DD2-135135-On-L, respectively.
The solid orange curves denote the result for the model DD2-135135-On-H-53, in which the average energy of the neutrino absorption reaction is estimated by Eq.~\eqref{eq:ave53} instead of Eq.~\eqref{eq:ave32}.
The green curves denote the results for the model DD2-135135-Iso-H, in which the the pair-annihilation heating is calculated assuming the isotropic angular distribution of neutrinos.
}
\label{fig:eje}
\end{center}
\end{figure}

We plot the results in Fig.~\ref{fig:eje}.
The top panel shows that the ejecta mass increases for $t \lesssim 100$ ms, but the growth rate decreases with time after that.
The ejecta mass reaches about $8\times 10^{-4}M_\odot$ finally.
The kinetic energy, shown in the middle panel, has the same trend as the ejecta mass, and its final value is about $1 \times 10^{49}$ erg.
These values are comparable to those of the dynamical ejecta for the models with the DD2 EOS, which are $\sim 10^{-3}M_\odot$ and $\sim2\times10^{49}$ erg \citep{2015PhRvD..91f4059S}.

Using these quantities, we estimate the average velocity of the ejecta $V_{\rm ej}$ by
\begin{align}
V_{\rm ej} = \sqrt{\frac{2E_{\rm kin,ej}}{M_{\rm b,ej}}}.
\end{align}
Note that we do not consider the contribution from a component that was already unbound at the beginning of the simulation, i.e., we only take into account the neutrino-driven ejecta.
We show the evolution of $V_{\rm ej}$ in the bottom panel of Fig.~\ref{fig:eje}.
The average velocity of the ejecta is initially about $0.18\ c$, but decreases to $0.13\ c$ in the quasi-stationary phase.
The initial enhancement for $V_{\rm ej}$ is induced by the strong pair-annihilation heating.

In Fig.~\ref{fig:eje}, we also plot the results for the model DD2-135135-Iso-H, in which the pair-annihilation heating rate is taken into account assuming the isotropic angular distribution for neutrinos.
The ejecta mass and the kinetic energy for the model DD2-135135-Iso-H are about 1.5 and 4 times larger than those for the fiducial model, respectively.
Hence, the averaged velocity of the ejecta for the model DD2-135135-Iso-H is larger as $\sim 0.2\ c$ than that in the fiducial model.
However, the relativistic outflow is not observed even in this model.
The reason would be that the specific heating rate is not sufficiently high even in this model.

\begin{figure}[t]
\begin{center}
\includegraphics[bb=0 0 360 216,width=\hsize]{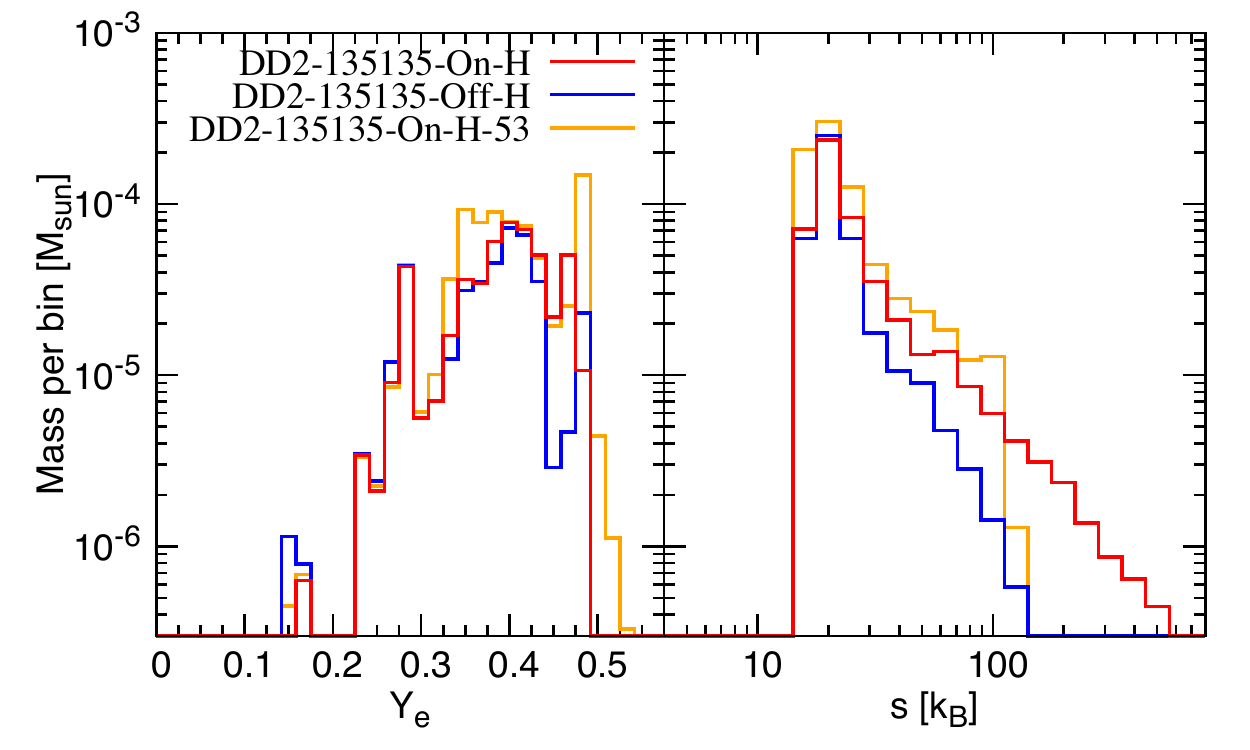}
\caption{
Mass histogram of the neutrino-driven ejecta at $t=300$ ms.
The distributions of the electron fraction (left panel) and the specific entropy (right panel) are shown.
The red solid curves denote the results for the fiducial model DD2-135135-On-H, and the blue dashed curves denote the results in the absence of the pair-annihilation heating (i.e., the model DD2-135135-Off-H).
The orange curves denote the results in which Eq.~\eqref{eq:ave53} is used for the neutrino absorption heating (i.e., the model DD2-135135-On-H-53).
}
\label{fig:hist}
\end{center}
\end{figure}

\begin{figure}[t]
\begin{center}
\includegraphics[bb=0 0 360 504,width=\hsize]{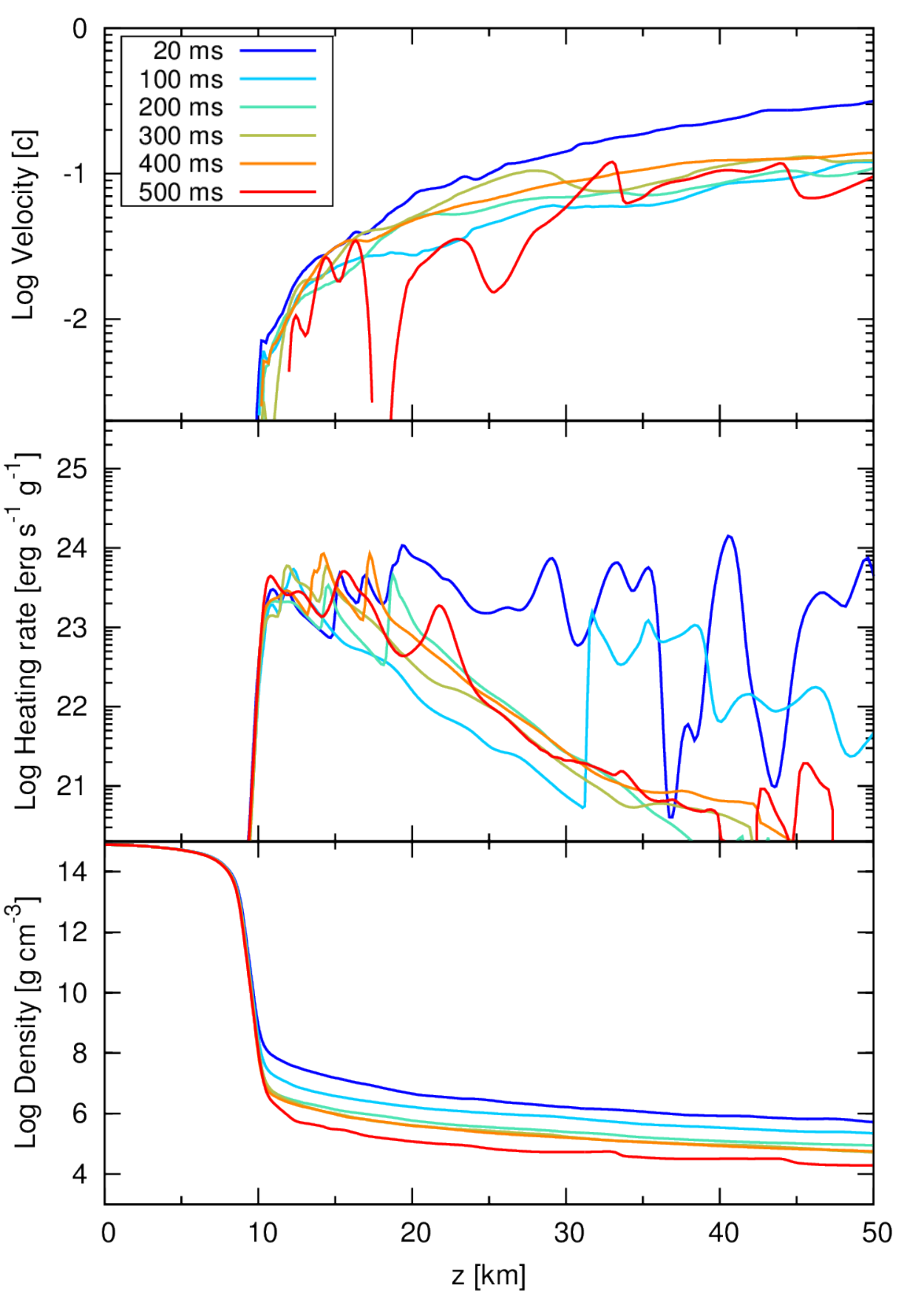}
\caption{
Profiles of the velocity (top), the specific heating rate due to the neutrino pair-annihilation (middle), and the rest-mass density (bottom) along the rotational axis at the timeslices  $t=$20, 100, 200, 300, 400, and 500 ms.
}
\label{fig:axrho}
\end{center}
\end{figure}

Next, we pay attention to the mass distributions of the electron fraction and the specific entropy of the ejecta.
Figure~\ref{fig:hist} shows the mass histogram of the ejecta $Y_e$ and entropy at $t=300$ ms, at which the ejecta mass becomes approximately constant in time (see Fig.~\ref{fig:eje}).
We find from the left panel that the ejecta are mildly neutron-rich, and have relatively high electron fraction $Y_e>0.25$, with the typical value $\sim 0.4$.
This property depends only weakly on the presence of the pair-annihilation heating.
This is because the fluid velocity just above the MNS ($z\approx 10$ km) is very small $v\sim 0.01\ c$ even in the presence of the pair-annihilation heating (see the top panel of Fig.~\ref{fig:axrho}).
Hence the neutrino absorption timescale in the polar region, which is estimated using the average energy of the neutrinos $\left<\omega\right>$ (in $\hbar=1$ unit) by
\begin{align}
&\frac{4\pi r^2 \left<\omega\right>}{L_\nu} \frac{1}{G_{\rm F}^2 \left<\omega\right>^2} \notag\\
&\sim 4\times 10^{-4}  \biggl(\frac{\left<\omega \right>}{10\ {\rm MeV}}\biggr)^{-1}\biggl(\frac{L_\nu}{10^{53}\ {\rm erg/s}}\biggr)^{-1}\biggl(\frac{r}{10\ {\rm km}}\biggr)^2\ {\rm s},
\end{align}
is shorter than the time for the fluid element needed to escape from the polar region
\begin{align}
z/v \sim 3\times 10^{-3}  \biggl(\frac{z}{10\ {\rm km}}\biggr)\biggl(\frac{v}{0.01c}\biggr)^{-1}\ {\rm s}. \label{eq:extime}
\end{align}
That is, the electron fraction of the ejecta achieves an equilibrium value \citep[e.g., see Eq.~(77) in][]{1996ApJ...471..331Q} soon after the ejecta is launched from the MNS.

From the right panel of Fig.~\ref{fig:hist}, we find that the typical specific entropy of the ejecta is $\sim 10\ k_{\rm B}$.
We also find that a small fraction of the ejecta has very high specific entropy; the highest value is $500\ k_{\rm B}$.
The pair-annihilation heating process generates the high-entropy ejecta because this process heats up the material regardless of the baryon density, and hence, a large amount of thermal energy can be injected in the low baryon density region.
Implication of these results on the nucleosynthesis is discussed in Sec.~\ref{sec:r}.

\subsection{Dependence of Ejecta Properties on Average-Energy Estimation of Neutrinos} \label{sec:energy}
Here we discuss the dependence of ejecta properties on the methods of estimating the average energy of neutrinos.
The orange curves in Figs. \ref{fig:eje} and \ref{fig:hist} denote the mass, kinetic energy, average velocity, and mass histogram of the ejecta for the model DD2-135135-On-H-53, in which the average energy is estimated by Eq. \eqref{eq:ave53}.
We find that the ejecta mass is enhanced, and that the high-entropy component of $s\gtrsim 200\ k_{\rm B}$ is absent for the model DD2-135135-On-H-53.

In the vicinity of the MNS and the torus, the neutrino absorption process dominates the heating rate due to the high density.
Thus, the increase of the neutrino absorption heating rate enhances the mass ejection from the system.
In general, $\sqrt{F_5(\eta_\nu)/F_3(\eta_\nu)}$ is larger than $F_3(\eta_\nu)/F_2(\eta_\nu)$, and hence, the neutrino absorption heating rate estimated by Eq.~\eqref{eq:ave53} are larger than those by Eq.~\eqref{eq:ave32} in the entire region.
As a result, the ejecta mass is also larger for the model DD2-135135-On-H-53 as found in Fig.~\ref{fig:eje}.

Because of the enhancement of the mass ejection from the MNS, the mass density around the rotational axis increases, and the specific heating rate decreases in the pole region where the pair-annihilation heating is efficient.
As a result, the highest end of the specific entropy of the ejecta for the model DD2-135135-On-H-53 does not become as large as that for the model DD2-135135-On-H.

The average energy is thus a key quantity for launching the relativistic outflow because the baryon-loading would be more significant if the neutrino absorption heating is larger.
We have to keep in mind that the condition for the relativistic outflow discussed in Sec \ref{sec:grb} will depend on the estimation of the average energy of neutrinos.

To rigorously discuss the average energy of neutrinos and the neutrino absorption heating rates, a simulation with energy-dependent radiation transfer scheme is needed.
Although such simulation is beyond scope of our present work, we have to perform it in the future in order to derive more convincing numerical results.

\subsection{Convergence Test}
In Figs.~\ref{fig:lumpair} and~\ref{fig:eje}, we plot the results for the lower resolution models DD2-135135-On-M (solid) and DD2-135135-On-L (dashed) by the thin curves.
For $t\lesssim 100$ ms, the time evolution of the neutrino luminosity depends weakly on the grid resolution.
The disagreement in the results among the three different resolution models is $\lesssim$5\ \% , so that a convergence is reasonably achieved during this time.
On the other hand, the convergence becomes poor for the late time.
The difference in the luminosity of electron antineutrinos and heavy-lepton neutrinos between the highest- and lowest-resolution models is $\sim$40\ \% at $t=400$ ms, while the luminosity of electron neutrinos does not depend on the resolution significantly.

The resolution dependence of the average energy of neutrinos shows the same trend as that of the luminosity.
The increases of the average energy of electron antineutrinos and heavy-lepton neutrinos in time become weaker in higher resolution models, and thus, we can only discuss the upper limit of the average energy of these neutrino species.
The possible reason for this behavior is that the density gradient at the surface of the MNS, around which neutrinos are most significantly emitted, becomes steeper at that time, and hence, the diffusion of the neutrino emission is not accurately resolved with the low resolution.
Taking this resolution dependence into account, we may conclude that $L_{\bar{\nu}_e}\lesssim 10^{52}\ {\rm erg\ s^{-1}}$, $L_{\nu_x}\lesssim 7\times 10^{51}\ {\rm erg\ s^{-1}}$, $\left<\omega_{\bar{\nu}_e}\right>_{\rm em} \lesssim 13\ {\rm MeV}$, and $\left<\omega_{\nu_x}\right>_{\rm em}\lesssim 20\ {\rm MeV}$ at $t=400$ ms.

For the ejected mass and its kinetic energy (see Fig.~\ref{fig:eje}), the convergence is better achieved.
The difference between the highest- and lowest-resolution models is $\sim$ 3\ \% and $\sim$ 2\ \% for the mass and the kinetic energy of the ejecta, respectively.

\section{Discussion}
\subsection{Can Neutrinos Drive SGRBs?} \label{sec:grb}
As mentioned in Sec.~\ref{sc:intro}, the neutrino pair-annihilation could be a driving force for launching a relativistic ejecta, which could be SGRBs.
In the results of our fiducial model (DD2-135135-On-H), the kinetic energy of the ejecta ($\sim 10^{49}\ {\rm erg}$) is smaller than the typical energy (the sum of the energy of gamma-rays in prompt emission and the kinetic energy of the blast wave) of SGRBs of $\sim 10^{50}\ {\rm erg}$ \citep{2015ApJ...815..102F}.
Moreover, no relativistic outflow is observed.
The primary reason for this is the presence of high-density baryon in the polar region of the MNS.
Even in the model in which the pair-annihilation heating is optimistically evaluated (DD2-135135-Iso-H), the kinetic energy of the ejecta ($\sim 4\times10^{49}\ {\rm erg}$) is still lower than $\sim 10^{50}\ {\rm erg}$, and the Lorentz factor of the ejecta is also small ($\Gamma\lesssim1.3$).

We can estimate the terminal Lorentz factor of the ejecta as
\begin{align}
\Gamma_{\rm f} = \frac{Q}{\rho} \tau_{\rm heat} \approx 1.1\biggl(\frac{Q/\rho}{\rm 10^{24}\ erg\ g^{-1}\ s^{-1}}\biggr)\biggl(\frac{\tau_{\rm heat}}{\rm 1\ ms}\biggr), \label{eq:gamma}
\end{align}
where $Q$ is the heating rate, and $\tau_{\rm heat}\sim r/v$ is the heating timescale defined by the length scale of the heating region divided by the fluid velocity of the region.
Here, as fiducial values, we take $Q/\rho \sim 10^{24}\ {\rm erg\ g^{-1}\ s^{-1}}$ and $\tau_{\rm heat} \sim 30\ {\rm km}/0.1\ c \approx 1$ ms from the results at $t=20$ ms shown in Fig.~\ref{fig:axrho}.
Note that we assumed that the internal energy deposited in the material due to heating processes would be totally transformed into the kinetic energy.
The small Lorentz factor in our simulation ($\Gamma\lesssim1.3$) is consistent with the estimate of Eq.~\eqref{eq:gamma}.

\begin{figure}[t]
\begin{center}
\includegraphics[bb=0 0 360 201,width=\hsize]{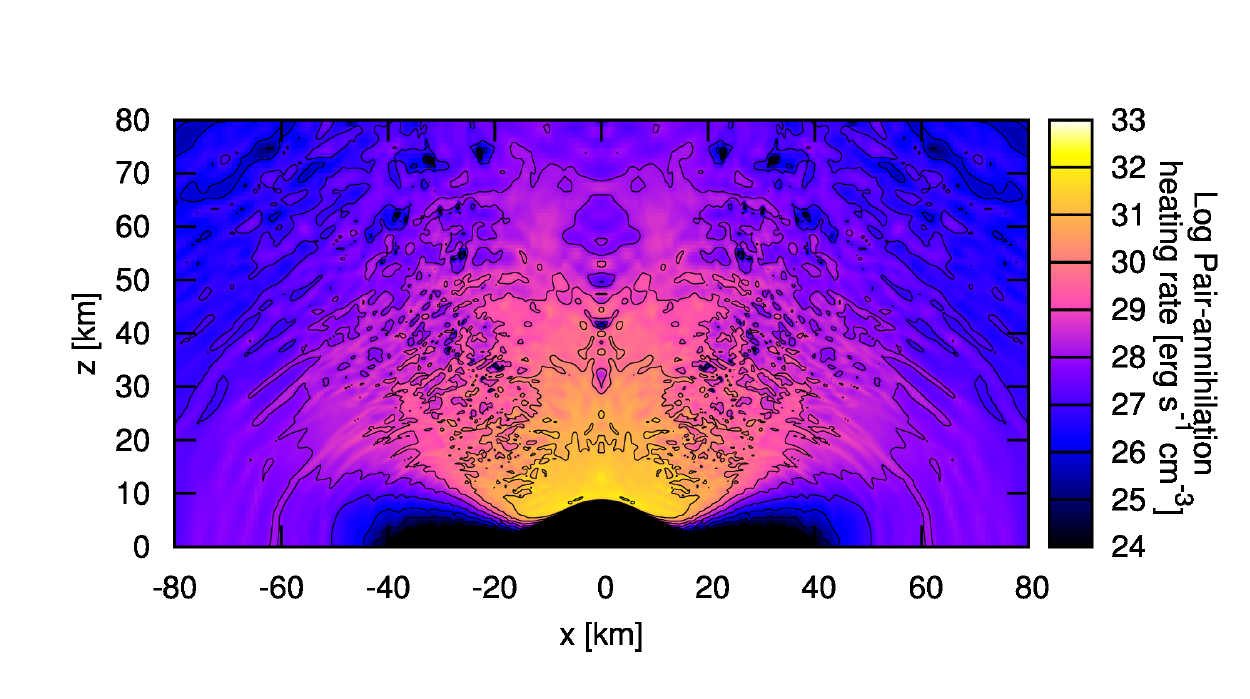}
\caption{
Snapshot of the neutrino pair-annihilation heating rate for the fiducial model DD2-135135-On-H at $t=$ 10 ms in the meridional plane.
The black curves denote logarithmically spaced contours with the intervals of 1.0 dex.
}
\label{fig:pr}
\end{center}
\end{figure}

The Lorentz factor of $\sim100$ could be obtained if the specific heating rate is $Q/\rho \sim 10^{26}\ {\rm erg\ g^{-1}\ s^{-1}}$ over a length scale of $\sim 30$ km.
We plot the density structure along the rotational axis in the bottom panel of Fig.~\ref{fig:axrho}.
This shows that the density decreases with time gradually, so that the heating rate required to achieve the outflow with the high Lorentz factor also decreases with time.
For example, for $t\gtrsim$400 ms, the density decreases to $\lesssim 10^5\ {\rm g\ cm^{-3}}$.
Then the heating rate required to launch the ultra-relativistic ejecta is $Q\sim 10^{31}\ {\rm erg\ cm^{-3}\ s^{-1}}$.
Such a high heating rate is indeed achieved in the first $\sim 10$ ms in our simulation (see Fig.~\ref{fig:pr}), at which the system has very large neutrino luminosity of $\sim 10^{53}\ {\rm erg\ s^{-1}}$.
Therefore, the relativistic outflow could be launched for later times if such a high luminosity could be sustained.

\subsection{Nucleosynthesis in the Neutrino-driven Ejecta} \label{sec:r}
Most of the neutrino-driven ejecta found in this paper have the electron fraction between 0.3 and 0.5 and the specific entropy $\sim 10-20\ k_{\rm B}$.
Therefore, the strong $r$-process may not occur in this type of the ejecta \citep{1997ApJ...482..951H,2015ApJ...813....2M}.

One interesting finding is that, due to the effect of the pair-annihilation heating, a small amount of very high specific entropy material of $s\sim 500\ k_{\rm B}$ is ejected (see Fig.~\ref{fig:hist}).
The electron fraction of such ejecta is $\sim0.45-0.50$, and the expansion velocity is $\sim 0.2-0.5\ c$.
In such high-entropy and fast-expanding material, heavy nuclei could be synthesized through the $r$-process \citep[the condition for the $r$-process nucleosynthesis is found, e.g, see ][]{1997ApJ...482..951H}.
Even in a slightly proton-rich condition, if the entropy and expansion velocity of the ejecta are sufficiently high, a lot of alpha particles and nucleons remain in low-temperature environment due to alpha-rich freeze-out.
Then the heavy nuclei could be produced through the nucleon capture process as described in \cite{2002PhRvL..89w1101M} and \cite{2016ApJ...818...96F}.
To explore these issues, we need a detailed nucleosynthesis calculation.

\subsection{Effects of the Viscosity} \label{sec:vis}
In our simulations, the viscous effect, which is likely to be induced by the magnetohydrodynamical turbulent motion \citep{1998RvMP...70....1B,2013ApJ...772..102H,2014ApJ...784..121S,2016MNRAS.457..857S,2016MNRAS.456.2273S} is not taken into account.
Recent general relativistic magnetohydrodynamics simulations for NS-NS mergers suggest that strongly magnetized MNS and torus are likely to be formed after the merger via the Kelvin-Helmholtz and the magnetorotational instabilities \citep{2014PhRvD..90d1502K,2015PhRvD..92l4034K}.
The turbulent flow in the torus is likely to be sustained by the magnetorotational instability, and then, the flow will transport the angular momentum.
Furthermore, the dissipation of the turbulent motion will heat up the torus.
The dynamics of the system could be changed significantly by the effective viscosity.

First, the properties of the torus could be modified.
For example, the geometrical thickness of the torus could be increased due to the viscous heating, and in addition, the torus could spread outward due to the angular momentum transport \citep{2017arXiv170506142S}.
Furthermore, the viscous heating could rise the temperature inside the torus, and as a result, the neutrino luminosity of the torus would be increased.
Then the heating due to neutrinos would be enhanced and will help driving a relativistic jet from the polar region of MNS as discussed in Sec. \ref{sec:grb}.

Second, in addition to the neutrino-driven wind, the viscosity-driven wind from the torus is expected.
As suggested in the previous works \citep{2013MNRAS.435..502F,2014MNRAS.441.3444M,2015MNRAS.446..750F,2015PhRvD..92f4034K,2017arXiv170505473S}, $\sim 10$ \% of the torus mass may be ejected as the viscosity-driven wind, so that the properties of the mass ejection in the viscous time scale $\sim 100$ ms would be affected by the viscous effect.
We plan to perform viscous radiation hydrodynamics simulations in our next work.

\begin{figure}[t]
\begin{center}
\includegraphics[bb=0 0 360 216,width=\hsize]{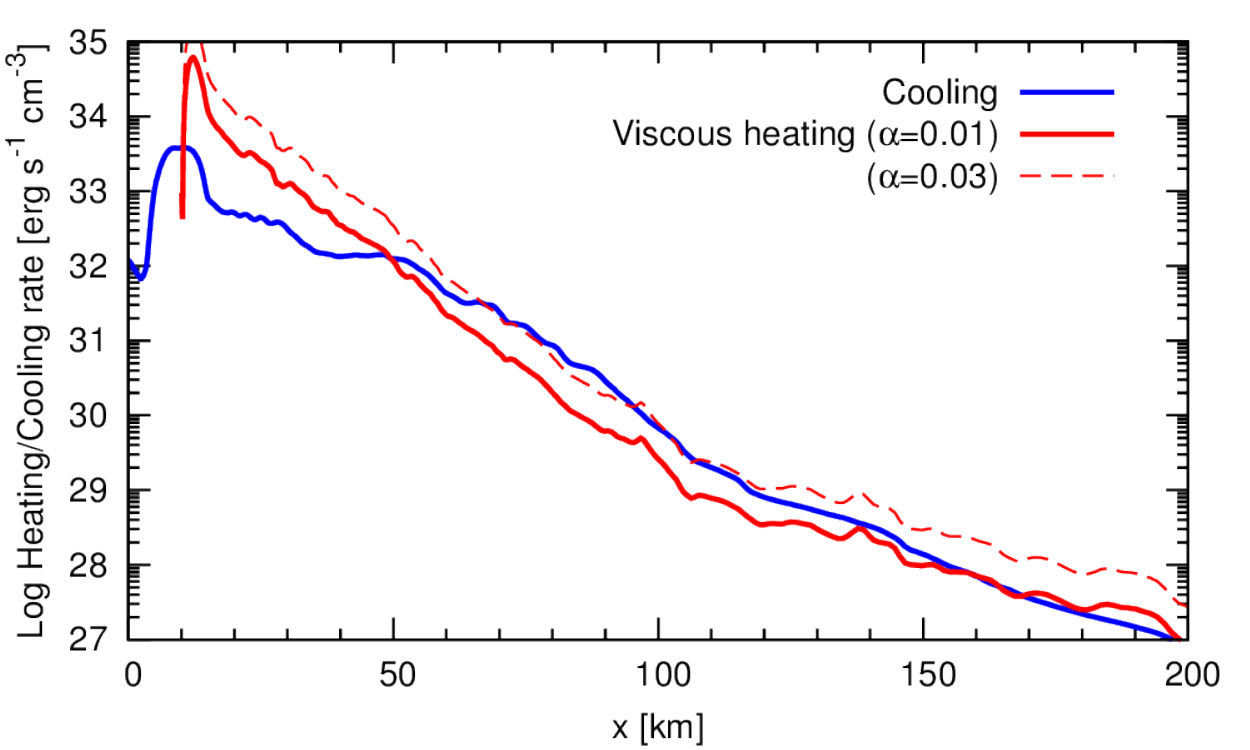}
\caption{
Viscous heating rates (red curves) and neutrino emission cooling rate (blue curve).
For the viscous heating rates, we assume $\alpha=0.01$ for the solid curve and $\alpha=0.03$ for the dashed curve.
}
\label{fig:vtime}
\end{center}
\end{figure}

As the prelude, we roughly estimate the viscous effects using a snapshot of the present non-viscous simulations.
In an axisymmetric, geometrically thin, and stationary accretion disk, the viscous heating rate $Q^{(+)}_{\rm vis}$ becomes
\begin{align}
Q^{(+)}_{\rm vis}=\rho \nu \biggl(x\frac{\del \Omega}{\del x}\biggr)^2,
\end{align}
where we employed a Shakura-Sunyaev parametrization of the kinetic viscosity coefficient \citep{1973A&A....24..337S,2008bhad.book.....K} as
\begin{align}
\nu = \alpha \frac{c_s^2}{\Omega} = \alpha \frac{c_s^2 x}{v_\varphi}.
\end{align}
We show the viscous heating rate and neutrino emission cooling rate ($\sum_i Q_{({\rm leak})\nu_i}$) on the equatorial plane as functions of the radius in Fig.~\ref{fig:vtime}.
Here we used the axisymmetric configuration at the beginning of the simulation.
We assumed $\alpha\sim 10^{-2}$ as a fiducial value following the latest results by high-resolution magnetohydrodynamics simulations \citep{2013ApJ...772..102H,2014ApJ...784..121S,2016MNRAS.457..857S,2016MNRAS.456.2273S}.
We find that the viscous heating rate is comparable to the cooling rate by the neutrino emission.
This suggests that, due to the viscous heating, the energy loss by the neutrino cooling may be compensated, and the temperature of the torus would decrease more slowly than that in the simulation without the viscosity.
Therefore, in reality, the initially large neutrino luminosity could be sustained for a longer timescale and the mass and the kinetic energy of the neutrino-driven ejecta would be larger in the presence of the viscous effect.

\subsection{Uncertainties of Pair-annihilation Heating Rate}\label{subsec:vol-int}
It is known that the M1-closure scheme has a problem in the optically thin region \citep[e.g.,][]{2014ApJS..213....7J,2016ApJ...818..162O,2016ApJ...826...23T,2006JCoPh.218....1F}.
As a result, the pair-annihilation heating rate evaluated in our present work can be significantly different from more realistic one.
This can be understood, for example, from the fact that the cross section of the pair-annihilation process has strong angle-dependence ($\propto (1-\cos \Theta)^2$ with $\Theta$ being the scattering angle between two neutrinos).

In \cite{2015MNRAS.448..541J}, radiation fields obtained with their M1 scheme are compared to those obtained with a ray-tracing method.
It is shown that the radiation energy density around the rotational axis is unphysically enhanced in the M1 scheme.
Although they compared only the energy density and flux of neutrinos, their result implies that our pair-annihilation heating rate could be overestimated because of the enhancement of energy density.
On the other hand, our estimation of the average energy deposited per reaction (see Eq.~\eqref{eq:ave}) could be underestimated because we use the local matter temperature for its estimation.

Here, we estimate the heating rates in terms of a leakage-based volume-integration method described in \cite{1997A&A...319..122R} using each snapshot of the simulation and compare the heating rate with that of our M1-scheme.
According to \cite{1997A&A...319..122R}, the heating rate due to the pair-annihilation in this scheme can be written (in $\hbar=1$ unit) as
\begin{widetext}
\begin{align}
Q^{(+)}_{{\rm pair},\nu_i} = \frac{C_{\nu_i \bar{\nu}_i}^{\rm pair} G_{\rm F}^2}{3\pi} \int dV_1\int dV_2 \left<\omega\right>_{\rm pair} Q^{(-)}_{\rm (leak)}{}_{\nu_i}({\bm x}_1) Q^{(-)}_{\rm (leak)}{}_{\bar{\nu}_i}({\bm x}_2) \frac{(1-\cos\Theta)^2}{\pi^2|{\bm x}-{\bm x}_1||{\bm x}-{\bm x}_2|},
\end{align}
\end{widetext}
where $dV=d^3x\sqrt{\gamma}$ is the physical volume element, $\cos\Theta = (x-x_1)\cdot(x-x_2)/|x-x_1||x-x_2|$ is the angle between the momenta of colliding neutrinos, and $Q^{(-)}_{\rm (leak)}{}_{\nu_i}(x_1)$ is the leakage source term for neutrinos of the flavor $i$ except for the heating processes.
Here we do not consider general relativistic effects such as the bending of neutrino trajectories and gravitational redshift.
For the average energy of neutrinos $\left<\omega\right>_{\rm pair}$, we employ the value used in the radiation hydrodynamics simulation (i.e., Eq.~\eqref{eq:ave}) and focus only on the difference in the scattering angle.

\begin{figure}[t]
\begin{center}
\includegraphics[bb=0 0 360 216,width=\hsize]{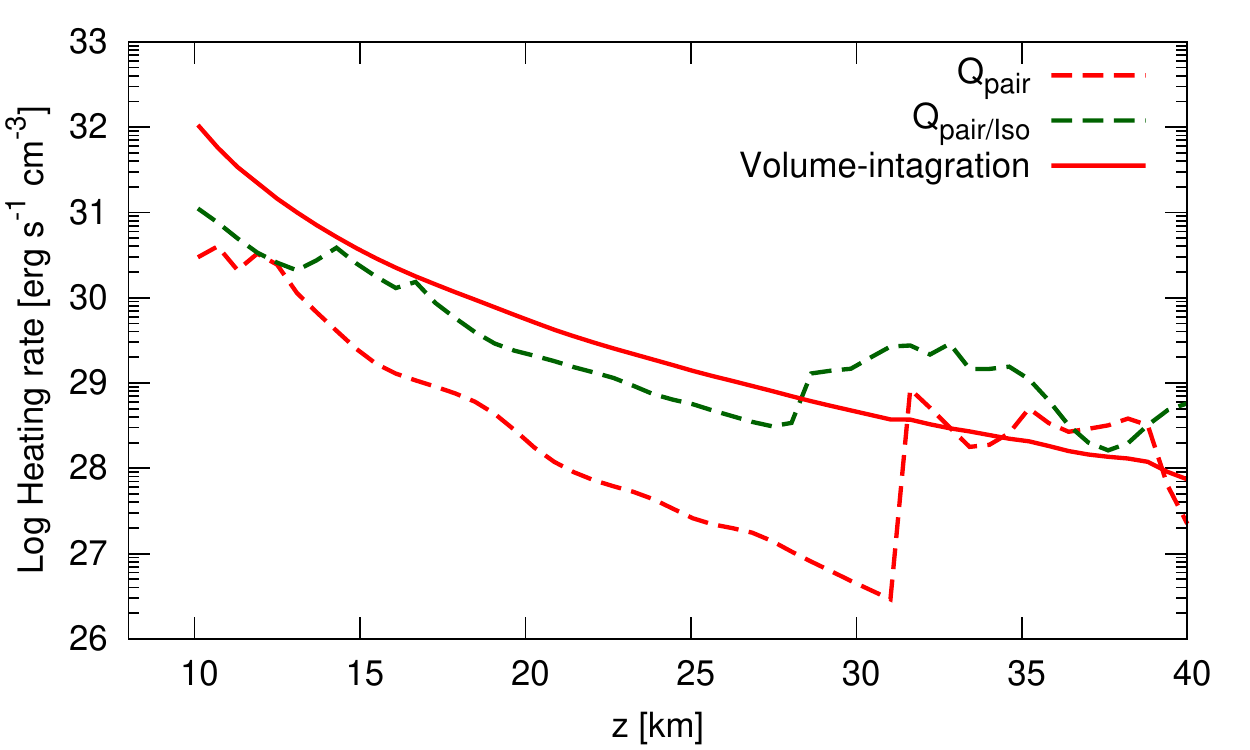}
\caption{
Neutrino pair-annihilation heating rate along the rotational axis at $t=100$ ms.
The red solid curve denotes the heating rate calculated with the volume-integration method.
The red and green dashed curves denote the heating rates for the models DD2-135135-On-H and DD2-135135-Iso-H, respectively.
}
\label{fig:vol-int}
\end{center}
\end{figure}

Figure~\ref{fig:vol-int} shows the pair-annihilation heating rate using the volume-integration method along the rotational ($z$-) axis.
It is found that the heating rate based on the moment formalism is by a factor of $\gtrsim 10$ smaller than the results in the volume-integration framework.

On the other hand, the pair-annihilation heating rate based on the assumption of isotropic angular distribution of neutrinos could provide a result that agrees with that in the volume-integration framework in a better manner than that in the moment formalism for the region shown in Fig.~\ref{fig:vol-int}.
This is because, near the MNS, neutrinos propagate in various directions due to the neutrino emission from the torus and the MNS.
This suggests that, for the region at which the pair-annihilation heating is efficient, we need to take into account the angular distribution of neutrinos more carefully.

\section{Summary}
We performed a fully general relativistic, axisymmetric numerical relativity simulation for a MNS surrounded by a torus, which is the typical remnant of the NS-NS merger.
We took into account neutrino transport using the truncated moment formalism with a M1-closure and relevant weak interaction reactions between neutrinos and the fluid material including the neutrino pair-annihilation in an approximate manner.
For the initial condition of this simulation, we used a configuration obtained in a three-dimensional, numerical relativity simulation for the NS-NS merger.
Our purpose is to investigate the amount and the properties of the material ejected due to the neutrino heating in the framework of purely radiation-hydrodynamics simulation.

In this setup, a quasi-stationary neutrino-driven outflow is launched for $\sim$300 ms from the beginning of the simulation.
The effect of the neutrino pair-annihilation heating is quite large because of the very high neutrino luminosity of the MNS and the torus.
Furthermore, due to the existence of the dense and hot torus, the structure of the heating rate density is quite different from the isolated NS usually considered as normal core-collapse supernova remnants.
For the DD2 EOS, the mass and the kinetic energy of the neutrino-driven ejecta are comparable to those of the dynamical ejecta \citep{2015PhRvD..91f4059S}.
We expect that this would be also the case for stiff EOS or for the merger of low-mass NS.
Therefore, the neutrino-driven ejecta would contribute to the mass and kinetic energy of the entire ejecta of the NS-NS merger for this EOS.

The relativistic outflow required for SGRBs is not found in our present simulation because the specific heating rate around the rotational axis is not sufficiently high for driving it.
The reasons for this might be the small pair-annihilation heating rate and the baryon pollution due to the neutrino-driven wind from the MNS.
Moreover, the kinetic energy of the ejecta is smaller than the typical value of the SGRBs.
Our results suggest that, in this purely-radiation hydrodynamics case, the neutrino pair-annihilation process in MNS-torus systems cannot account for the majority of observed SGRB events.
However, we used an approximate neutrino transport scheme at this point, and hence the final conclusion should be drawn by performing simulations fully solving Boltzmann's equation for neutrino transport.
Moreover, we do not consider the effects of magnetohydrodynamic turbulence in our simulations although the MNS is likely to be strongly magnetized and the resulting turbulent motion could play an important role for its evolution \citep{2015PhRvD..92l4034K}.
Hence the neutrino emissivity could be underestimated because the viscous heating associated with the turbulence motion would enhance the neutrino luminosity.
For this reason, in the future, we plan to perform more sophisticated simulations considering these missing elements.

In order to obtain the relativistic outflow from this MNS-torus system, the high neutrino luminosity of $\sim 10^{53}\ {\rm erg\ s^{-1}}$ would be needed at later time, for which the density in the polar region decreases to $\lesssim 10^{5}\ {\rm g\ cm^{-3}}$.
Such high neutrino luminosity may be achieved due to the viscous heating in the accretion torus.

\acknowledgements
S. F. thanks Kunihito Ioka, Yudai Suwa, and Takahiro Tanaka for fruitful discussions, Kohsuke Sumiyoshi for discussions about the radiation transfer and the emission processes of neutrinos, and the anonymous referee for helping to improve the quality of this article.
This work was initiated during S.F.'s stay in Toho University, and S. F. also thanks Tetsu Kitayama for willingly accepting S.F.'s visit to the university.
S. F. is supported by Research Fellowship of Japan Society for the Promotion of Science (JSPS) for Young Scientists (No.26-1329). 
This work was supported by JSPS Grant-in-Aid for Scientific Research (16H02183, 15H00836, 15K05077, 15H00783, 16K17706, 16H06341, 15H00782), by the HPCI Strategic Program of Japanese MEXT, and by a MEXT program ``Priority Issue 9 to be tackled by using Post K computer".
Numerical simulations were carried out on Cray XC40 in Yukawa Institute for Theoretical Physics, Kyoto University and FX10 in Information Technology Center, the University of Tokyo.
Numerical analyses were carried out on computers at Center for Computational Astrophysics, National Astronomical Observatory of Japan.

\bibliographystyle{apj}
\bibliography{apj-jour,reference}

\begin{thebibliography}{95}
\expandafter\ifx\csname natexlab\endcsname\relax\def\natexlab#1{#1}\fi

\bibitem[{{Abadie} {et~al.}(2010){Abadie}, {Abbott}, {Abbott}, {Abernathy},
  {Adams}, {Adhikari}, {Ajith}, {Allen}, {Allen}, {Amador Ceron}, \&
  et~al.}]{2010NIMPA.624..223A}
{Abadie}, J., {et~al.} 2010, Nuclear Instruments and Methods in Physics
  Research A, 624, 223

\bibitem[{{Accadia} {et~al.}(2011){Accadia}, {Acernese}, {Antonucci}, {Astone},
  {Ballardin}, {Barone}, {Barsuglia}, {Basti}, {Bauer}, {Bebronne}, {Beker},
  {Belletoile}, {Birindelli}, {Bitossi}, {Bizouard}, {Blom}, {Bondu},
  {Bonelli}, {Bonnand}, {Boschi}, {Bosi}, {Bouhou}, {Braccini}, {Bradaschia},
  {Branchesi}, {Briant}, {Brillet}, {Brisson}, {Budzy{\'n}ski}, {Bulik},
  {Bulten}, {Buskulic}, {Buy}, {Cagnoli}, {Calloni}, {Canuel}, {Carbognani},
  {Cavalier}, {Cavalieri}, {Cella}, {Cesarini}, {Chaibi}, {Chassande Mottin},
  {Chincarini}, {Cleva}, {Coccia}, {Cohadon}, {Colacino}, {Colas}, {Colla},
  {Colombini}, {Corsi}, {Coulon}, {Cuoco}, {D'Antonio}, {Dattilo}, {Davier},
  {Day}, {De Rosa}, {Debreczeni}, {Del Pozzo}, {del Prete}, {Di Fiore}, {Di
  Lieto}, {Emilio}, {Di Virgilio}, {Dietz}, {Drago}, {Fafone}, {Ferrante},
  {Fidecaro}, {Fiori}, {Flaminio}, {Forte}, {Fournier}, {Franc}, {Frasca},
  {Frasconi}, {Galimberti}, {Gammaitoni}, {Garufi}, {G{\'a}sp{\'a}r}, {Gemme},
  {Genin}, {Gennai}, {Giazotto}, {Gouaty}, {Granata}, {Greverie}, {Guidi},
  {Hayau}, {Heidmann}, {Heitmann}, {Hello}, {Huet}, {Jaranowski}, {Kowalska},
  {Kr{\'o}lak}, {Leroy}, {Letendre}, {Li}, {Liguori}, {Lorenzini}, {Loriette},
  {Losurdo}, {Majorana}, {Maksimovic}, {Man}, {Mantovani}, {Marchesoni},
  {Marion}, {Marque}, {Martelli}, {Masserot}, {Michel}, {Milano}, {Minenkov},
  {Mohan}, {Morgado}, {Morgia}, {Mosca}, {Moscatelli}, {Mours}, {Nocera},
  {Pagliaroli}, {Palladino}, {Palomba}, {Paoletti}, {Parisi}, {Pasqualetti},
  {Passaquieti}, {Passuello}, {Persichetti}, {Pichot}, {Piergiovanni},
  {Pietka}, {Pinard}, {Poggiani}, {Prato}, {Prodi}, {Punturo}, {Puppo},
  {Rabeling}, {R{\'a}cz}, {Rapagnani}, {Re}, {Regimbau}, {Ricci}, {Robinet},
  {Rocchi}, {Rolland}, {Romano}, {Rosi{\'n}ska}, {Ruggi}, {Sassolas},
  {Sentenac}, {Sperandio}, {Sturani}, {Swinkels}, {Tacca}, {Taffarello},
  {Toncelli}, {Tonelli}, {Torre}, {Tournefier}, {Travasso}, {Vajente}, {van den
  Brand}, {Van Den Broeck}, {van der Putten}, {Vasuth}, {Vavoulidis},
  {Vedovato}, {Verkindt}, {Vetrano}, {Vicer{\'e}}, {Vinet}, {Vitale}, {Vocca},
  {Ward}, {Was}, {Yvert}, \& {Zendri}}]{2011CQGra..28k4002A}
{Accadia}, T., {et~al.} 2011, Classical and Quantum Gravity, 28, 114002

\bibitem[{{Alcubierre} {et~al.}(2003){Alcubierre}, {Br{\"u}gmann}, {Diener},
  {Koppitz}, {Pollney}, {Seidel}, \& {Takahashi}}]{2003PhRvD..67h4023A}
{Alcubierre}, M., {Br{\"u}gmann}, B., {Diener}, P., {Koppitz}, M., {Pollney},
  D., {Seidel}, E., \& {Takahashi}, R. 2003, \prd, 67, 084023

\bibitem[{{Alcubierre} {et~al.}(2001){Alcubierre}, {Br{\"u}gmann}, {Holz},
  {Takahashi}, {Brandt}, {Seidel}, {Thornburg}, \&
  {Ashtekar}}]{2001IJMPD..10..273A}
{Alcubierre}, M., {Br{\"u}gmann}, B., {Holz}, D., {Takahashi}, R., {Brandt},
  S., {Seidel}, E., {Thornburg}, J., \& {Ashtekar}, A. 2001, International
  Journal of Modern Physics D, 10, 273

\bibitem[{{Antoniadis} {et~al.}(2013){Antoniadis}, {Freire}, {Wex}, {Tauris},
  {Lynch}, {van Kerkwijk}, {Kramer}, {Bassa}, {Dhillon}, {Driebe}, {Hessels},
  {Kaspi}, {Kondratiev}, {Langer}, {Marsh}, {McLaughlin}, {Pennucci}, {Ransom},
  {Stairs}, {van Leeuwen}, {Verbiest}, \& {Whelan}}]{2013Sci...340..448A}
{Antoniadis}, J., {et~al.} 2013, Science, 340, 448

\bibitem[{{Baker} {et~al.}(2006){Baker}, {Centrella}, {Choi}, {Koppitz}, \&
  {van Meter}}]{2006PhRvL..96k1102B}
{Baker}, J.~G., {Centrella}, J., {Choi}, D.-I., {Koppitz}, M., \& {van Meter},
  J. 2006, Physical Review Letters, 96, 111102

\bibitem[{{Balbus} \& {Hawley}(1998)}]{1998RvMP...70....1B}
{Balbus}, S.~A., \& {Hawley}, J.~F. 1998, Reviews of Modern Physics, 70, 1

\bibitem[{{Banik} {et~al.}(2014){Banik}, {Hempel}, \&
  {Bandyopadhyay}}]{2014ApJS..214...22B}
{Banik}, S., {Hempel}, M., \& {Bandyopadhyay}, D. 2014, \apjs, 214, 22

\bibitem[{{Baumgarte} \& {Shapiro}(1999)}]{1999PhRvD..59b4007B}
{Baumgarte}, T.~W., \& {Shapiro}, S.~L. 1999, \prd, 59, 024007

\bibitem[{{Blandford} \& {Znajek}(1977)}]{1977MNRAS.179..433B}
{Blandford}, R.~D., \& {Znajek}, R.~L. 1977, \mnras, 179, 433

\bibitem[{{Br{\"u}gmann} {et~al.}(2008){Br{\"u}gmann}, {Gonz{\'a}lez},
  {Hannam}, {Husa}, {Sperhake}, \& {Tichy}}]{2008PhRvD..77b4027B}
{Br{\"u}gmann}, B., {Gonz{\'a}lez}, J.~A., {Hannam}, M., {Husa}, S.,
  {Sperhake}, U., \& {Tichy}, W. 2008, \prd, 77, 024027

\bibitem[{{Burrows} {et~al.}(2006){Burrows}, {Reddy}, \&
  {Thompson}}]{2006NuPhA.777..356B}
{Burrows}, A., {Reddy}, S., \& {Thompson}, T.~A. 2006, Nuclear Physics A, 777,
  356

\bibitem[{{Campanelli} {et~al.}(2006){Campanelli}, {Lousto}, {Marronetti}, \&
  {Zlochower}}]{2006PhRvL..96k1101C}
{Campanelli}, M., {Lousto}, C.~O., {Marronetti}, P., \& {Zlochower}, Y. 2006,
  Physical Review Letters, 96, 111101

\bibitem[{{Cooperstein}(1988)}]{1988PhR...163...95C}
{Cooperstein}, J. 1988, \physrep, 163, 95

\bibitem[{{Cooperstein} {et~al.}(1986){Cooperstein}, {van den Horn}, \&
  {Baron}}]{1986ApJ...309..653C}
{Cooperstein}, J., {van den Horn}, L.~J., \& {Baron}, E.~A. 1986, \apj, 309,
  653

\bibitem[{{Demorest} {et~al.}(2010){Demorest}, {Pennucci}, {Ransom}, {Roberts},
  \& {Hessels}}]{2010Natur.467.1081D}
{Demorest}, P.~B., {Pennucci}, T., {Ransom}, S.~M., {Roberts}, M.~S.~E., \&
  {Hessels}, J.~W.~T. 2010, \nat, 467, 1081

\bibitem[{{Dessart} {et~al.}(2009){Dessart}, {Ott}, {Burrows}, {Rosswog}, \&
  {Livne}}]{2009ApJ...690.1681D}
{Dessart}, L., {Ott}, C.~D., {Burrows}, A., {Rosswog}, S., \& {Livne}, E. 2009,
  \apj, 690, 1681

\bibitem[{{Dietrich} {et~al.}(2015){Dietrich}, {Bernuzzi}, {Ujevic}, \&
  {Br{\"u}gmann}}]{2015PhRvD..91l4041D}
{Dietrich}, T., {Bernuzzi}, S., {Ujevic}, M., \& {Br{\"u}gmann}, B. 2015, \prd,
  91, 124041

\bibitem[{{Eichler} {et~al.}(1989){Eichler}, {Livio}, {Piran}, \&
  {Schramm}}]{1989Natur.340..126E}
{Eichler}, D., {Livio}, M., {Piran}, T., \& {Schramm}, D.~N. 1989, \nat, 340,
  126

\bibitem[{{Fernandez} {et~al.}(2013){Fernandez}, {Mani}, {Rinaldi}, {Kadau},
  {Mosquet}, {Lombois-Burger}, {Cayer-Barrioz}, {Herrmann}, {Spencer}, \&
  {Isa}}]{2013PhRvL.111j8301F}
{Fernandez}, N., {et~al.} 2013, Physical Review Letters, 111, 108301

\bibitem[{{Fern{\'a}ndez} {et~al.}(2015){Fern{\'a}ndez}, {Kasen}, {Metzger}, \&
  {Quataert}}]{2015MNRAS.446..750F}
{Fern{\'a}ndez}, R., {Kasen}, D., {Metzger}, B.~D., \& {Quataert}, E. 2015,
  \mnras, 446, 750

\bibitem[{{Fern{\'a}ndez} \& {Metzger}(2013)}]{2013MNRAS.435..502F}
{Fern{\'a}ndez}, R., \& {Metzger}, B.~D. 2013, \mnras, 435, 502

\bibitem[{{Fischer} {et~al.}(2010){Fischer}, {Whitehouse}, {Mezzacappa},
  {Thielemann}, \& {Liebend{\"o}rfer}}]{2010A&A...517A..80F}
{Fischer}, T., {Whitehouse}, S.~C., {Mezzacappa}, A., {Thielemann}, F.-K., \&
  {Liebend{\"o}rfer}, M. 2010, \aap, 517, A80

\bibitem[{{Fong} {et~al.}(2015){Fong}, {Berger}, {Margutti}, \&
  {Zauderer}}]{2015ApJ...815..102F}
{Fong}, W., {Berger}, E., {Margutti}, R., \& {Zauderer}, B.~A. 2015, \apj, 815,
  102

\bibitem[{{Foucart} {et~al.}(2016){Foucart}, {Haas}, {Duez}, {O'Connor}, {Ott},
  {Roberts}, {Kidder}, {Lippuner}, {Pfeiffer}, \&
  {Scheel}}]{2016PhRvD..93d4019F}
{Foucart}, F., {et~al.} 2016, \prd, 93, 044019

\bibitem[{{Frank} {et~al.}(2006){Frank}, {Dubroca}, \&
  {Klar}}]{2006JCoPh.218....1F}
{Frank}, M., {Dubroca}, B., \& {Klar}, A. 2006, Journal of Computational
  Physics, 218, 1

\bibitem[{{Freiburghaus} {et~al.}(1999){Freiburghaus}, {Rosswog}, \&
  {Thielemann}}]{1999ApJ...525L.121F}
{Freiburghaus}, C., {Rosswog}, S., \& {Thielemann}, F.-K. 1999, \apjl, 525,
  L121

\bibitem[{{Fujibayashi} {et~al.}(2016){Fujibayashi}, {Yoshida}, \&
  {Sekiguchi}}]{2016ApJ...818...96F}
{Fujibayashi}, S., {Yoshida}, T., \& {Sekiguchi}, Y. 2016, \apj, 818, 96

\bibitem[{{Fuller} {et~al.}(1985){Fuller}, {Fowler}, \&
  {Newman}}]{1985ApJ...293....1F}
{Fuller}, G.~M., {Fowler}, W.~A., \& {Newman}, M.~J. 1985, \apj, 293, 1

\bibitem[{{Gonz{\'a}lez} {et~al.}(2007){Gonz{\'a}lez}, {Audit}, \&
  {Huynh}}]{2007A&A...464..429G}
{Gonz{\'a}lez}, M., {Audit}, E., \& {Huynh}, P. 2007, \aap, 464, 429

\bibitem[{{Goriely} {et~al.}(2011){Goriely}, {Bauswein}, \&
  {Janka}}]{2011ApJ...738L..32G}
{Goriely}, S., {Bauswein}, A., \& {Janka}, H.-T. 2011, \apjl, 738, L32

\bibitem[{{Hawley} {et~al.}(2013){Hawley}, {Richers}, {Guan}, \&
  {Krolik}}]{2013ApJ...772..102H}
{Hawley}, J.~F., {Richers}, S.~A., {Guan}, X., \& {Krolik}, J.~H. 2013, \apj,
  772, 102

\bibitem[{{Hempel} {et~al.}(2012){Hempel}, {Fischer}, {Schaffner-Bielich}, \&
  {Liebend{\"o}rfer}}]{2012ApJ...748...70H}
{Hempel}, M., {Fischer}, T., {Schaffner-Bielich}, J., \& {Liebend{\"o}rfer}, M.
  2012, \apj, 748, 70

\bibitem[{{Hoffman} {et~al.}(1997){Hoffman}, {Woosley}, \&
  {Qian}}]{1997ApJ...482..951H}
{Hoffman}, R.~D., {Woosley}, S.~E., \& {Qian}, Y.-Z. 1997, \apj, 482, 951

\bibitem[{{Hotokezaka} {et~al.}(2013){Hotokezaka}, {Kiuchi}, {Kyutoku},
  {Muranushi}, {Sekiguchi}, {Shibata}, \& {Taniguchi}}]{2013PhRvD..88d4026H}
{Hotokezaka}, K., {Kiuchi}, K., {Kyutoku}, K., {Muranushi}, T., {Sekiguchi},
  Y.-i., {Shibata}, M., \& {Taniguchi}, K. 2013, \prd, 88, 044026

\bibitem[{{Hotokezaka} {et~al.}(2011){Hotokezaka}, {Kyutoku}, {Okawa},
  {Shibata}, \& {Kiuchi}}]{2011PhRvD..83l4008H}
{Hotokezaka}, K., {Kyutoku}, K., {Okawa}, H., {Shibata}, M., \& {Kiuchi}, K.
  2011, \prd, 83, 124008

\bibitem[{{Jiang} {et~al.}(2014){Jiang}, {Stone}, \&
  {Davis}}]{2014ApJS..213....7J}
{Jiang}, Y.-F., {Stone}, J.~M., \& {Davis}, S.~W. 2014, \apjs, 213, 7

\bibitem[{{Just} {et~al.}(2015){Just}, {Bauswein}, {Pulpillo}, {Goriely}, \&
  {Janka}}]{2015MNRAS.448..541J}
{Just}, O., {Bauswein}, A., {Pulpillo}, R.~A., {Goriely}, S., \& {Janka}, H.-T.
  2015, \mnras, 448, 541

\bibitem[{{Just} {et~al.}(2016){Just}, {Obergaulinger}, {Janka}, {Bauswein}, \&
  {Schwarz}}]{2016ApJ...816L..30J}
{Just}, O., {Obergaulinger}, M., {Janka}, H.-T., {Bauswein}, A., \& {Schwarz},
  N. 2016, \apjl, 816, L30

\bibitem[{{Kaplan} {et~al.}(2014){Kaplan}, {Ott}, {O'Connor}, {Kiuchi},
  {Roberts}, \& {Duez}}]{2014ApJ...790...19K}
{Kaplan}, J.~D., {Ott}, C.~D., {O'Connor}, E.~P., {Kiuchi}, K., {Roberts}, L.,
  \& {Duez}, M. 2014, \apj, 790, 19

\bibitem[{{Kasen} {et~al.}(2013){Kasen}, {Badnell}, \&
  {Barnes}}]{2013ApJ...774...25K}
{Kasen}, D., {Badnell}, N.~R., \& {Barnes}, J. 2013, \apj, 774, 25

\bibitem[{{Kato} {et~al.}(2008){Kato}, {Fukue}, \&
  {Mineshige}}]{2008bhad.book.....K}
{Kato}, S., {Fukue}, J., \& {Mineshige}, S. 2008, {Black-Hole Accretion Disks
  --- Towards a New Paradigm ---} ({Kyoto University Press})

\bibitem[{{Kiuchi} {et~al.}(2015{\natexlab{a}}){Kiuchi}, {Cerd{\'a}-Dur{\'a}n},
  {Kyutoku}, {Sekiguchi}, \& {Shibata}}]{2015PhRvD..92l4034K}
{Kiuchi}, K., {Cerd{\'a}-Dur{\'a}n}, P., {Kyutoku}, K., {Sekiguchi}, Y., \&
  {Shibata}, M. 2015{\natexlab{a}}, \prd, 92, 124034

\bibitem[{{Kiuchi} {et~al.}(2014){Kiuchi}, {Kyutoku}, {Sekiguchi}, {Shibata},
  \& {Wada}}]{2014PhRvD..90d1502K}
{Kiuchi}, K., {Kyutoku}, K., {Sekiguchi}, Y., {Shibata}, M., \& {Wada}, T.
  2014, \prd, 90, 041502

\bibitem[{{Kiuchi} {et~al.}(2015{\natexlab{b}}){Kiuchi}, {Sekiguchi},
  {Kyutoku}, {Shibata}, {Taniguchi}, \& {Wada}}]{2015PhRvD..92f4034K}
{Kiuchi}, K., {Sekiguchi}, Y., {Kyutoku}, K., {Shibata}, M., {Taniguchi}, K.,
  \& {Wada}, T. 2015{\natexlab{b}}, \prd, 92, 064034

\bibitem[{{Kiuchi} {et~al.}(2009){Kiuchi}, {Sekiguchi}, {Shibata}, \&
  {Taniguchi}}]{2009PhRvD..80f4037K}
{Kiuchi}, K., {Sekiguchi}, Y., {Shibata}, M., \& {Taniguchi}, K. 2009, \prd,
  80, 064037

\bibitem[{{Kiuchi} {et~al.}(2010){Kiuchi}, {Sekiguchi}, {Shibata}, \&
  {Taniguchi}}]{2010PhRvL.104n1101K}
---. 2010, Physical Review Letters, 104, 141101

\bibitem[{{Kuroda} \& {LCGT Collaboration}(2010)}]{2010CQGra..27h4004K}
{Kuroda}, K., \& {LCGT Collaboration}. 2010, Classical and Quantum Gravity, 27,
  084004

\bibitem[{{Lattimer} \& {Schramm}(1974)}]{1974ApJ...192L.145L}
{Lattimer}, J.~M., \& {Schramm}, D.~N. 1974, \apjl, 192, L145

\bibitem[{{Levermore}(1984)}]{1984JQSRT..31..149L}
{Levermore}, C.~D. 1984, J. Quant. Spectrosc. Radiat. Transfer, 31, 149

\bibitem[{{Li} \& {Paczy{\'n}ski}(1998)}]{1998ApJ...507L..59L}
{Li}, L.-X., \& {Paczy{\'n}ski}, B. 1998, \apjl, 507, L59

\bibitem[{{Lippuner} {et~al.}(2017){Lippuner}, {Fern{\'a}ndez}, {Roberts},
  {Foucart}, {Kasen}, {Metzger}, \& {Ott}}]{2017arXiv170306216L}
{Lippuner}, J., {Fern{\'a}ndez}, R., {Roberts}, L.~F., {Foucart}, F., {Kasen},
  D., {Metzger}, B.~D., \& {Ott}, C.~D. 2017, ArXiv e-prints

\bibitem[{{Marronetti} {et~al.}(2008){Marronetti}, {Tichy}, {Br{\"u}gmann},
  {Gonz{\'a}lez}, \& {Sperhake}}]{2008PhRvD..77f4010M}
{Marronetti}, P., {Tichy}, W., {Br{\"u}gmann}, B., {Gonz{\'a}lez}, J., \&
  {Sperhake}, U. 2008, \prd, 77, 064010

\bibitem[{{Martin} {et~al.}(2015){Martin}, {Perego}, {Arcones}, {Thielemann},
  {Korobkin}, \& {Rosswog}}]{2015ApJ...813....2M}
{Martin}, D., {Perego}, A., {Arcones}, A., {Thielemann}, F.-K., {Korobkin}, O.,
  \& {Rosswog}, S. 2015, \apj, 813, 2

\bibitem[{{Meszaros} \& {Rees}(1992)}]{1992MNRAS.257P..29M}
{Meszaros}, P., \& {Rees}, M.~J. 1992, \mnras, 257, 29P

\bibitem[{{Metzger} \& {Fern{\'a}ndez}(2014)}]{2014MNRAS.441.3444M}
{Metzger}, B.~D., \& {Fern{\'a}ndez}, R. 2014, \mnras, 441, 3444

\bibitem[{{Meyer}(2002)}]{2002PhRvL..89w1101M}
{Meyer}, B.~S. 2002, Physical Review Letters, 89, 231101

\bibitem[{{Nakar}(2007)}]{2007PhR...442..166N}
{Nakar}, E. 2007, \physrep, 442, 166

\bibitem[{{Narayan} {et~al.}(1992){Narayan}, {Paczynski}, \&
  {Piran}}]{1992ApJ...395L..83N}
{Narayan}, R., {Paczynski}, B., \& {Piran}, T. 1992, \apjl, 395, L83

\bibitem[{{Ohsuga} \& {Takahashi}(2016)}]{2016ApJ...818..162O}
{Ohsuga}, K., \& {Takahashi}, H.~R. 2016, \apj, 818, 162

\bibitem[{{Perego} {et~al.}(2014){Perego}, {Rosswog}, {Cabez{\'o}n},
  {Korobkin}, {K{\"a}ppeli}, {Arcones}, \&
  {Liebend{\"o}rfer}}]{2014MNRAS.443.3134P}
{Perego}, A., {Rosswog}, S., {Cabez{\'o}n}, R.~M., {Korobkin}, O.,
  {K{\"a}ppeli}, R., {Arcones}, A., \& {Liebend{\"o}rfer}, M. 2014, \mnras,
  443, 3134

\bibitem[{{Perego} {et~al.}(2017){Perego}, {Yasin}, \&
  {Arcones}}]{2017arXiv170102017P}
{Perego}, A., {Yasin}, H., \& {Arcones}, A. 2017, ArXiv e-prints

\bibitem[{{Qian} \& {Woosley}(1996)}]{1996ApJ...471..331Q}
{Qian}, Y.-Z., \& {Woosley}, S.~E. 1996, \apj, 471, 331

\bibitem[{{Richers} {et~al.}(2015){Richers}, {Kasen}, {O'Connor},
  {Fern{\'a}ndez}, \& {Ott}}]{2015ApJ...813...38R}
{Richers}, S., {Kasen}, D., {O'Connor}, E., {Fern{\'a}ndez}, R., \& {Ott},
  C.~D. 2015, \apj, 813, 38

\bibitem[{{Rosswog} {et~al.}(1999){Rosswog}, {Liebend{\"o}rfer}, {Thielemann},
  {Davies}, {Benz}, \& {Piran}}]{1999A&A...341..499R}
{Rosswog}, S., {Liebend{\"o}rfer}, M., {Thielemann}, F.-K., {Davies}, M.~B.,
  {Benz}, W., \& {Piran}, T. 1999, \aap, 341, 499

\bibitem[{{Ruffert} {et~al.}(1996){Ruffert}, {Janka}, \&
  {Schaefer}}]{1996A&A...311..532R}
{Ruffert}, M., {Janka}, H.-T., \& {Schaefer}, G. 1996, \aap, 311, 532

\bibitem[{{Ruffert} {et~al.}(1997){Ruffert}, {Janka}, {Takahashi}, \&
  {Schaefer}}]{1997A&A...319..122R}
{Ruffert}, M., {Janka}, H.-T., {Takahashi}, K., \& {Schaefer}, G. 1997, \aap,
  319, 122

\bibitem[{{Salmonson} \& {Wilson}(1999)}]{1999ApJ...517..859S}
{Salmonson}, J.~D., \& {Wilson}, J.~R. 1999, \apj, 517, 859

\bibitem[{{Salvesen} {et~al.}(2016){Salvesen}, {Simon}, {Armitage}, \&
  {Begelman}}]{2016MNRAS.457..857S}
{Salvesen}, G., {Simon}, J.~B., {Armitage}, P.~J., \& {Begelman}, M.~C. 2016,
  \mnras, 457, 857

\bibitem[{{Sekiguchi} {et~al.}(2011){Sekiguchi}, {Kiuchi}, {Kyutoku}, \&
  {Shibata}}]{2011PhRvL.107e1102S}
{Sekiguchi}, Y., {Kiuchi}, K., {Kyutoku}, K., \& {Shibata}, M. 2011, Physical
  Review Letters, 107, 051102

\bibitem[{{Sekiguchi} {et~al.}(2012){Sekiguchi}, {Kiuchi}, {Kyutoku}, \&
  {Shibata}}]{2012PTEP.2012aA304S}
---. 2012, Progress of Theoretical and Experimental Physics, 2012, 01A304

\bibitem[{{Sekiguchi} {et~al.}(2015){Sekiguchi}, {Kiuchi}, {Kyutoku}, \&
  {Shibata}}]{2015PhRvD..91f4059S}
---. 2015, \prd, 91, 064059

\bibitem[{{Sekiguchi} {et~al.}(2016){Sekiguchi}, {Kiuchi}, {Kyutoku},
  {Shibata}, \& {Taniguchi}}]{2016PhRvD..93l4046S}
{Sekiguchi}, Y., {Kiuchi}, K., {Kyutoku}, K., {Shibata}, M., \& {Taniguchi}, K.
  2016, \prd, 93, 124046

\bibitem[{{Setiawan} {et~al.}(2004){Setiawan}, {Ruffert}, \&
  {Janka}}]{2004MNRAS.352..753S}
{Setiawan}, S., {Ruffert}, M., \& {Janka}, H.-T. 2004, \mnras, 352, 753

\bibitem[{{Shakura} \& {Sunyaev}(1973)}]{1973A&A....24..337S}
{Shakura}, N.~I., \& {Sunyaev}, R.~A. 1973, \aap, 24, 337

\bibitem[{{Shen} {et~al.}(1998{\natexlab{a}}){Shen}, {Toki}, {Oyamatsu}, \&
  {Sumiyoshi}}]{1998NuPhA.637..435S}
{Shen}, H., {Toki}, H., {Oyamatsu}, K., \& {Sumiyoshi}, K. 1998{\natexlab{a}},
  Nuclear Physics A, 637, 435

\bibitem[{{Shen} {et~al.}(1998{\natexlab{b}}){Shen}, {Toki}, {Oyamatsu}, \&
  {Sumiyoshi}}]{1998PThPh.100.1013S}
---. 1998{\natexlab{b}}, Progress of Theoretical Physics, 100, 1013

\bibitem[{{Shi} {et~al.}(2016){Shi}, {Stone}, \& {Huang}}]{2016MNRAS.456.2273S}
{Shi}, J.-M., {Stone}, J.~M., \& {Huang}, C.~X. 2016, \mnras, 456, 2273

\bibitem[{{Shibata}(2003{\natexlab{a}})}]{2003PhRvD..67b4033S}
{Shibata}, M. 2003{\natexlab{a}}, \prd, 67, 024033

\bibitem[{{Shibata}(2003{\natexlab{b}})}]{2003ApJ...595..992S}
---. 2003{\natexlab{b}}, \apj, 595, 992

\bibitem[{{Shibata} \& {Kiuchi}(2017)}]{2017arXiv170506142S}
{Shibata}, M., \& {Kiuchi}, K. 2017, \prd, 95, 123003

\bibitem[{{Shibata} {et~al.}(2011){Shibata}, {Kiuchi}, {Sekiguchi}, \&
  {Suwa}}]{2011PThPh.125.1255S}
{Shibata}, M., {Kiuchi}, K., {Sekiguchi}, Y., \& {Suwa}, Y. 2011, Progress of
  Theoretical Physics, 125, 1255

\bibitem[{{Shibata} \& {Nakamura}(1995)}]{1995PhRvD..52.5428S}
{Shibata}, M., \& {Nakamura}, T. 1995, \prd, 52, 5428

\bibitem[{{Shibata} \& {Taniguchi}(2006)}]{2006PhRvD..73f4027S}
{Shibata}, M., \& {Taniguchi}, K. 2006, \prd, 73, 064027

\bibitem[{{Shibata} {et~al.}(2005){Shibata}, {Taniguchi}, \& {Ury{\=
  u}}}]{2005PhRvD..71h4021S}
{Shibata}, M., {Taniguchi}, K., \& {Ury{\= u}}, K. 2005, \prd, 71, 084021

\bibitem[{{Siegel} \& {Metzger}(2017)}]{2017arXiv170505473S}
{Siegel}, D.~M., \& {Metzger}, B.~D. 2017, ArXiv e-prints

\bibitem[{{Suzuki} \& {Inutsuka}(2014)}]{2014ApJ...784..121S}
{Suzuki}, T.~K., \& {Inutsuka}, S.-i. 2014, \apj, 784, 121

\bibitem[{{Symbalisty} \& {Schramm}(1982)}]{1982ApL....22..143S}
{Symbalisty}, E., \& {Schramm}, D.~N. 1982, \aplett, 22, 143

\bibitem[{{Takahashi} {et~al.}(2016){Takahashi}, {Ohsuga}, {Kawashima}, \&
  {Sekiguchi}}]{2016ApJ...826...23T}
{Takahashi}, H.~R., {Ohsuga}, K., {Kawashima}, T., \& {Sekiguchi}, Y. 2016,
  \apj, 826, 23

\bibitem[{{Takami} {et~al.}(2015){Takami}, {Rezzolla}, \&
  {Baiotti}}]{2015PhRvD..91f4001T}
{Takami}, K., {Rezzolla}, L., \& {Baiotti}, L. 2015, \prd, 91, 064001

\bibitem[{{Tanaka} \& {Hotokezaka}(2013)}]{2013ApJ...775..113T}
{Tanaka}, M., \& {Hotokezaka}, K. 2013, \apj, 775, 113

\bibitem[{{Thorne}(1981)}]{1981MNRAS.194..439T}
{Thorne}, K.~S. 1981, \mnras, 194, 439

\bibitem[{{Timmes} \& {Swesty}(2000)}]{2000ApJS..126..501T}
{Timmes}, F.~X., \& {Swesty}, F.~D. 2000, \apjs, 126, 501

\bibitem[{{Usov}(1992)}]{1992Natur.357..472U}
{Usov}, V.~V. 1992, \nat, 357, 472

\bibitem[{{Wanajo} {et~al.}(2014){Wanajo}, {Sekiguchi}, {Nishimura}, {Kiuchi},
  {Kyutoku}, \& {Shibata}}]{2014ApJ...789L..39W}
{Wanajo}, S., {Sekiguchi}, Y., {Nishimura}, N., {Kiuchi}, K., {Kyutoku}, K., \&
  {Shibata}, M. 2014, \apjl, 789, L39

\end{thebibliography}

\appendix
\section{A. Prescription of Neutrino-antineutrino Pair-annihilation Heating Rate in Energy-integrated M1-scheme} \label{app:pair}
In the fluid rest frame, the energy deposition rate for a neutrino species $\nu_i$ due to the pair-annihilation process is written (in $\hbar=1$ unit) as
\begin{align}
Q_{{\rm pair},\nu_i}^{(+)} = \int \frac{d^3k}{(2\pi)^3}\int \frac{d^3k'}{(2\pi)^3} \omega f_{\nu_i}(k) f_{\bar{\nu}_i}(k') \sigma_{{\rm pair},i} (k,k'), \label{eq:pair}
\end{align}
where $f_{\nu_i}(k)$ and $f_{\bar{\nu}_i}(k')$ are the distribution functions of the $i$-th species of neutrinos and antineutrinos, which are functions of momenta $k$ and $k'$, respectively, and $\omega=-u^\alpha k_\alpha$ is the energy of neutrinos in the fluid rest frame.
The cross section of the pair-annihilation process ($\nu_i+\bar{\nu}_i \rightarrow e^-+e^+$), $\sigma_{{\rm pair},i} (k,k')$, is written \citep[see, e.g.,][]{1999ApJ...517..859S} as
\begin{align}
\sigma_{{\rm pair},i} (k,k') \approx  \frac{C_{\nu_i \bar{\nu}_i}^{\rm pair} G_{\rm F}^2}{3 \pi}\frac{(k\cdot k')^2}{\omega \omega'} = \frac{C_{\nu_i \bar{\nu}_i}^{\rm pair} G_{\rm F}^2}{3 \pi} \omega \omega' (1-\ell^\alpha \ell'_\alpha)^2,
\end{align}
where we ignored the phase space blocking for electrons and positrons, and the electron mass ($\approx 0.511$ MeV) because the energy of neutrinos considered in this paper ($\agt 5$ MeV) is much higher.
Here we defined a spatial unit four-vector $\ell^\alpha$ orthogonal to $u^\alpha$ as
\begin{align}
k^\alpha = \omega (u^\alpha + \ell^\alpha).
\end{align}
In our moment formalism \citep{2011PThPh.125.1255S}, we define the energy-integrated zeroth-, first-, and second-rank moments of neutrinos by
\begin{align}
J&= \int d\omega d\Omega \biggl(\frac{\omega}{2\pi}\biggr)^3 f(k),\\
H^\alpha&= \int d\omega d\Omega  \biggl(\frac{\omega}{2\pi}\biggr)^3 f(k)\ell^\alpha,\\
L^{\alpha\beta}&= \int d\omega d\Omega  \biggl(\frac{\omega}{2\pi}\biggr)^3 f(k)\ell^\alpha \ell^\beta,
\end{align}
where $\Omega$ denotes the solid angle in the momentum space of neutrinos in the fluid rest frame.
In order to evaluate the heating rate using the above moments, we approximate the energy of neutrinos, $\omega$, by its average value $\left<\omega\right>_{\rm pair}$.
Then Eq.~\eqref{eq:pair} becomes
\begin{align}
Q_{{\rm pair},\nu_i}^{(+)} &\approx \left<\omega\right>_{\rm pair}  \frac{C_{\nu_i \bar{\nu}_i}^{\rm pair} G_{\rm F}^2}{3 \pi} \int d\omega d\Omega \biggl(\frac{\omega}{2\pi}\biggr)^3 \int d\omega' d\Omega' \biggl(\frac{\omega'}{2\pi}\biggr)^3 f_{\nu_i}(k) f_{\bar{\nu}_i}(k') (1-\ell^\alpha \ell'_\alpha)^2 \notag \\
&=\frac{C_{\nu_i \bar{\nu}_i}^{\rm pair} G_{\rm F}^2}{3\pi} \left<\omega\right>_{\rm pair} \bigl(J \bar{J}-2H^\beta \bar{H}_\beta+L^{\beta\gamma} \bar{L}_{\beta\gamma}\bigr), \label{eq:pair2}
\end{align}
which appeared in Eq.~\eqref{eq:pairrate}.

\section{B. Prescription of Neutrino Absorption Processes in Energy-integrated M1-scheme} \label{app:abs}
In the same way as Eq. \eqref{eq:pair}, the source terms due to the absorption of electron (anti)neutrinos are written (in $\hbar=1$ unit) as
\begin{align}
Q_{{\rm abs},\nu_e/\bar{\nu}_e}^{(+)}{}^\alpha &= \int \frac{d^3k}{(2\pi)^3} k^\alpha f_{\nu_e/\bar{\nu}_e}(k) \sigma_{{\rm abs},\nu_e/\bar{\nu}_e} (k)\ n_{n/p}\notag\\
 &= \int d\omega d\Omega \biggl(\frac{\omega}{2\pi}\biggr)^3 (u^\alpha + \ell^\alpha) f_{\nu_e/\bar{\nu}_e}(k) \sigma_{{\rm abs},\nu_e/\bar{\nu}_e} (k)\ \frac{\rho X_{n/p}}{m_u}. \label{eq:abs}
\end{align}
The cross section of the absorption processes is written as
\begin{align}
\sigma_{{\rm abs},\nu_e/\bar{\nu}_e} (k) \approx \frac{(1+3g_{\rm A}^2)G_{\rm F}^2}{\pi} \omega^2,
\end{align}
where we ignored the electron mass and the difference between proton and neutron masses ($m_n-m_p\approx 1.293$ MeV).
As in Eq.~\eqref{eq:pair2}, the energy deposition rates are evaluated in our scheme as
\begin{align}
Q_{{\rm abs},\nu_e}^{(+)}{}^\alpha &\approx \frac{(1+3g_{\rm A}^2)G_{\rm F}^2}{\pi} \frac{\rho X_n}{m_u}\left<\omega_{\nu_e} \right>_{\rm abs}^2 (Ju^\alpha + H^\alpha),\\
Q_{{\rm abs},\bar{\nu}_e}^{(+)}{}^\alpha &\approx \frac{(1+3g_{\rm A}^2)G_{\rm F}^2}{\pi}\frac{\rho X_p}{m_u} \left<\omega_{\bar{\nu}_e} \right>_{\rm abs}^2 (\bar{J}u^\alpha + \bar{H}^\alpha).
\end{align}
The average energy estimated by Eq.~\eqref{eq:ave32} is the simple spectral average assuming the Fermi-Dirac-type energy distribution for neutrinos.
On the other hand, the average energy estimated by Eq.~\eqref{eq:ave53} is derived when we take the energy dependence of the absorption reaction into account.
We can directly perform the energy integral in Eq.~\eqref{eq:abs} as
\begin{align}
\int d\omega \biggl(\frac{\omega}{2\pi}\biggr)^3 f_{\nu}(k) \sigma_{{\rm abs},\nu_e/\bar{\nu}_e}(k) &= \int d\omega \biggl(\frac{\omega}{2\pi}\biggr)^3 \frac{1}{e^{\omega/T_\nu-\eta_\nu}+1} \frac{(1+3g_{\rm A}^2)G_{\rm F}^2}{\pi} \omega^2\notag\\
&= \frac{(1+3g_{\rm A}^2)G_{\rm F}^2}{\pi} \frac{T_\nu{}^6}{(2\pi)^3} F_5 (\eta)\notag\\
&= \frac{(1+3g_{\rm A}^2)G_{\rm F}^2}{\pi} \biggl[T_\nu{}^2 \frac{F_5 (\eta_\nu)}{F_3(\eta_\nu)} \biggr]\frac{J}{4\pi},
\end{align}
where we used Eq.~\eqref{eq:Jetanu}.
Thus, the average energy of this reaction is the square root of the square bracket, i.e., Eq.~\eqref{eq:ave53}.

Using these rates, the source terms for electrons (Eq.\eqref{eq:ye}) are written as
\begin{align}
{\cal R}_{{\rm abs},{\nu_e}} &\approx \frac{Q_{{\rm abs},{\nu_e}}^{(+)}{}^\alpha (-u_\alpha)}{\left<\omega_{{\nu_e}} \right>_{\rm abs}} \approx \frac{(1+3g_{\rm A}^2)G_{\rm F}^2}{\pi} \frac{\rho X_n}{m_u}\left<\omega_{{\nu_e}} \right>_{\rm abs} J,\\
{\cal R}_{{\rm abs},\bar{\nu}_e} &\approx \frac{Q_{{\rm abs},\bar{\nu}_e}^{(+)}{}^\alpha (-u_\alpha)}{\left<\omega_{\bar{\nu}_e} \right>_{\rm abs}} \approx \frac{(1+3g_{\rm A}^2)G_{\rm F}^2}{\pi}\frac{\rho X_p}{m_u} \left<\omega_{\bar{\nu}_e} \right>_{\rm abs} \bar{J}.
\end{align}

\end{document}